\documentclass[12pt,draftclsnofoot,onecolumn]{IEEEtran}
\ifCLASSINFOpdf

\else

\fi
\usepackage{algorithm} 		% format of the algorithm
\usepackage{algorithmic} 	% format of the algorithm
\usepackage{bm}
\usepackage{graphicx}
\usepackage{amsmath}
\usepackage{flushend}
\usepackage{cite}
\usepackage{amssymb}
\usepackage{multirow}

\usepackage{changebar}
\usepackage{epstopdf}
\usepackage{exscale}
\usepackage{relsize}
\usepackage{color}
\usepackage{graphicx}%------------------------------added by ZY
\usepackage{subfigure}
\usepackage{geometry}
\begin{document}
%
% paper title
% Titles are generally capitalized except for words such as a, an, and, as,
% at, but, by, for, in, nor, of, on, or, the, to and up, which are usually
% not capitalized unless they are the first or last word of the title.
% Linebreaks \\ can be used within to get better formatting as desired.
% Do not put math or special symbols in the title.
\title{Design of Reconfigurable Intelligent Surface-Aided Cross-Media Communications}

% author names and affiliations
% transmag papers use the long conference author name format.

\author{Mingming Wu,~\IEEEmembership{}
        Yue Xiao,~\IEEEmembership{Member, IEEE,}
        Yulan Gao,~\IEEEmembership{}
        and
        Ming Xiao,~\IEEEmembership{Senior Member, IEEE}
\thanks{

M. Wu, Y. Xiao and Y. Gao are with the National Key Laboratory of Science and Technology on Communications, University of Electronic Science and Technology of China, Chengdu 611731, China. The corresponding author is Y. Xiao (e-mail: mingwu@std.uestc.edu.cn, xiaoyue@uestc.edu.cn, yulangaomath@163.com).

M. Xiao is with the Department of Information Science and Engineering, Royal Institute of Technology (KTH) 10044 Stockholm, Sweden (e-mail: mingx@kth.se).
}
}

% The paper headers
\markboth{}%
{}
\maketitle

%\IEEEtitleabstractindextext{%
\begin{abstract}
A novel reconfigurable intelligent surface (RIS)-aided hybrid reflection/transmitter design is proposed for achieving information exchange in cross-media communications. In pursuit of the balance between energy efficiency and low-cost implementations, the cloud-management transmission protocol is adopted in the integrated multi-media system. Specifically, the messages of devices using heterogeneous propagation media, are firstly transmitted to the medium-matched AP, with the aid of the RIS-based dual-hop transmission. After the operation of intermediate frequency conversion, the access point (AP) uploads the received signals to the cloud for further demodulating and decoding process. Based on time division multiple access (TDMA), the cloud is able to distinguish the downlink data transmitted to different devices and transforms them into the input of the RIS controller via the dedicated control channel. Thereby, the RIS can passively reflect the incident carrier back into the original receiver with the exchanged information during the preallocated slots, following the idea of an index modulation-based transmitter. Moreover, the iterative optimization algorithm is utilized for optimizing the RIS phase, transmit rate and time allocation jointly in the delay-constrained cross-media communication model. Our simulation results demonstrate that the proposed RIS-based scheme can improve the end-to-end throughput than that of the AP-based transmission, the equal time allocation, the random and the discrete phase adjustment benchmarks.
\end{abstract}

% Note that keywords are not normally used for peerreview papers.
\begin{IEEEkeywords}
Cross-media communications, RIS.
\end{IEEEkeywords}

\section{Introduction}
The historic evolution of wireless communications from the first-generation (1G) to the fifth-generation (5G) systems mostly relies on the further exploitation of the radio frequency (RF) resource, increasing from 300MHz to sub-6GHz \cite{AndrewsJG-2014,AgiwalM-2016,AlsharifMH-2017}. Nevertheless, due to the explosive growth of data traffic in the foreseeable future \cite{Cisco-2020}, the scarce spectrum in the RF band may not support the emerging high-rate information services well in the sixth-generation (6G) wireless systems \cite{ZhangZ-2019,DangS-2019,SaadW-2020}. For the sake of improving the transmit rate, researchers turned attention to the higher electromagnetic domain with broadband characteristics, such as millimeter wave (mmwave), terahertz (THz) \cite{Boul3A-2018,RapTS-2019} and visible light bands \cite{MatheLEM-2019}.
In addition to electromagnetic waves, the potential for utilizing multiple propagation media is further considered, including acoustic waves \cite{Oceanic-2019}, biochemical molecules \cite{AkanOB-2017}, power lines \cite{Sharma-2017}, quantum communications \cite{BotsiP-2019}, aiming at adapting the features of multi-media to the requirements of the large-scale heterogeneous 6G networks more flexibly and extensively \cite{ZhangZ-2019,YangP-2019}.
The co-existence of multi-band (generally called multi-media) systems will boost the quality of service (QoS) in wireless communications, while posing new challenges for hardware design, resource management and network architecture, owing to the inherent differences among multi-media.

The basic concept of cross-media communications was proposed for dealing with the issue of integrating multi-media systems in an efficient and reliable manner \cite{LiY-2020}. By contrast to the available device-to-device communications in single medium systems, the process of medium conversion is required for information exchange between devices using heterogeneous media. According to the location of implementing medium conversion, cross-media approaches are mainly classified into access point (AP)-integrated, user-centric and cloud management.
As a straightforward solution, integrating multi-media modules into an AP dispenses with extra adjustments of current network architectures. For example, in the satellite-terrestrial integrated networks, the intelligent backhaul node (iBN) was equipped with both satellite and terrestrial modems for cross-band communications \cite{ArtigaX-2018}. However, the integrated AP compromises with the flexibility of network deployment and weakens the robustness against fading and shadow effects owing to the coupling of multi-media processing.
Thanks to the advancements of smart devices, terminals equipped with multi-media modems can be charged with the task of medium conversion. Hence, the multi-media APs can be geographically separated and users can freely switch access among heterogeneous medium-based APs, referred to be user-centric. For example, the authors of \cite{KashefM-2016} proposed a heterogeneous network structure composed of the separated RF and the visible light communications (VLC) APs, where terminals with RF/VLC modems could choose to receive data from either of the two APs. In competition with the AP-integrated one, the user-centric scheme obtains better network flexibility at the cost of imposing additional burdens on hardware design and resource consumption of terminals.
Naturally, supposing medium conversion is carried out via the third party except APs and terminals, the balance between the flexibility of networks and the low cost of devices may be achieved.

Following this idea, combining software defined network (SDN) and cloud radio access network (C-RAN) \cite{JainR-2013}, the cloud-management scheme is expected to realize information exchange in the cloud without the aid of any integrated AP and multi-media terminal. In an integrated mmWave and sub-6G wireless network, the authors in \cite{SemiariO-2019} explored two integrated schemes at the media access control (MAC) and packet data convergence protocol (PDCP) layers, depending on whether mmWave and sub-6G radio interfaces are located at the same or different APs, i.e., the AP-integrated and non-integrated scheme. In comparison to the integrated AP scheme without the need for the middlehaul link, the separated deployment of multi-media radio interfaces requires extra delay and signaling overheads for information exchange between radio interfaces. To address this issue, a semi-tight integration scheme is proposed, which separates the function of radio access in the PHY layer but integrates the function of information exchange in the upper layers for different media. From the perspective of hardware framework, semi-tight integration-based multi-media radio access networks deploy separated active antenna units (AAUs) for independent media access but share the same distributed unit (DU) and centralized unit (CU) for integrated data processing. By further connecting CUs to the cloud via the backhaul link, cross-media communications can be achieved between devices located in the same or different cells meanwhile the disadvantages of the non-integrated AP can be mitigated. So far few papers investigated the issue of cross-media communications in the context of cloud-management, which motivates our work in this paper.

Meanwhile, recently reconfigurable intelligent surface (RIS) enjoys growing attention in 6G and beyond wireless communications, as a brand-new technique capable of controlling the propagation environment intentionally and deterministically \cite{BasarE-2019,WuQ-2020}. Specifically, RIS consists of multiple programmable electromagnetic reflectors, each of which offers a reconfigurable phase independently. This allows RIS to make adaptive response to incident signals. Using the digital-to-analogue converter (DAC) to control the bias voltage on reflectors, RIS can adjust its phase to enhance multipath gains or mitigate interferences at the objective receiver.
In the absence of the line-of-sight (LOS) path, RIS can support a dual-hop communication scenario by adapting its reflection coefficients to channel coefficients, identical to the operation of precoding/beamforming in multiple-input multiple-output (MIMO) systems \cite{GuoH-2020,PengZ-2021}. For example in \cite{GuoH-2020}, the phase of RIS and the precoding matrix of the base station are jointly optimized, in order to maximize the weighted sum rate in multiuser MISO systems. The authors of \cite{PengZ-2021} employed RIS for full-duplex two-way communications between MIMO base station and multiple SISO users. On the other hand, RIS can also act as an index modulation-based transmitter with low-complexity and energy-efficient implementations \cite{BasarE-2020}. The authors of \cite{TangW-2019} showed that the modulation order of an RIS-aided transmitter could be extended to 8PSK base on a practical testbed.
Combined with massive MIMO techniques, holographic MIMO surfaces (HMIMOS)-enabled communications proposed in \cite{HuangC-2020} have attracted extensive attention.
Moreover, the integration of different frequency bands in 5G promotes further research into the RIS-based wideband and multiband systems \cite{LiT-2018,ZhangN-2020,CaiW-2022}. In \cite{LiT-2018}, the dual-band metasurface was designed for two unit elements of S- and K-bands to work in the same aperture simultaneously, which could be achieved by embedding the subcell at K-band in the metasurface at S-band. Considering the differences in the EM response of RIS among different frequency bands, the authors of \cite{CaiW-2022} investigated the practical reflection model of the RIS-assisted multiband systems.
However, apart from the application of RIS in the single- and multi-media systems such as sub-6 GHz/mmWave \cite{DaiL-2020}, mmWave/THz \cite{NieS-2019,Manjappa-2018}, VLC \cite{ValaC-2019}, RF/FSO \cite{NajafiM-2021}, its potential in dealing with cross-media communications still remains elusive.

Against this background, in this paper, we conceive an RIS-aided hybrid reflection/transmitter scheme for cross-media communications in the context of cloud-management, aiming at improving the end-to-end throughput of information exchange with energy-efficient and low-cost implementations. The main contributions are summarized as follows.

\begin{itemize}

\item To achieve cross-media information exchange, an RIS-aided hybrid reflection/transmitter structure supporting full-duplex transmission is presented in a dual-hop communication scenario, where the RIS not only assists terminals in adaptively reflecting signals towards the AP in the uplink, but also passively transmits the exchanged messages to the original receiver in the downlink. Based on the cloud-management protocol, the uplink and downlink received signals are expressed and the corresponding channel capacity is derived.

\item Upon the designed frame structure and the transmission protocol, we construct a delay-constrained cross-media information exchange model. Specifically, the RIS phase, transmit rate and time allocation are jointly optimized based on the max-min end-to-end throughput. Due to the significant differences of channel fading, bandwidth and transmit power among heterogenous media, the effect of rate discrepancy on the system performance has to be considered, leading to the additional delay constraints.

\item The heterogenous characteristics of different media, the non-convexity of the objective function and the coupling of different optimization variables pose new challenges on the mentioned optimization model. To deal with these issues, we firstly decouple the original problem into several subproblems, where the non-convexity of the RIS phase optimization is dealt with by the equivalent transformation and the approximate treatment. Then, two iterative optimization algorithms are proposed for the cases with and without delay constraints, respectively. Numerical results demonstrate that the proposed scheme can improve the end-to-end throughput than that of other alternative schemes, while keeping the feedback delay within a tolerant level for real-time cross-media communications.
\end{itemize}

The rest of this paper is organized as follows. In Section II, we describe the system model of RIS-aided cross-media communications and formulate the problem of information exchange as a max-min optimization model, involving the joint optimization of the RIS phase, transmit rate and time allocation. Section III presents the detailed process of the proposed algorithms in the cases with and without delay constrains. Numerical results are given in Section IV. Section V concludes our discourse.

\emph{Notation:} Scalars, vectors and matrices are denoted by lower case, boldface lower
and boldface upper case letters, respectively. For any matrix (vector) $\bm{M}$, ${\bm{M}^{\rm{T}}}$, ${\bm{M}^{\rm{H}}}$ and ${\bm{M}^{ - 1}}$ represent the transpose, the conjugate transpose and matrix inversion, respectively, while ${m^*}$ is the conjugate of the single element $m$. In addition, $\left|  \cdot  \right|$ , ${\rm{Tr}}( \cdot )$ and ${\rm{Re}}( \cdot )$ represent the operation of obtaining the 2-norm, trace and real part, respectively. Moreover, we use  ${\rm{E}}( \cdot )$ for the expectation and $ \odot $ to denote the dot product of two matrixes. Furthermore, for a vector $\bm{v}$, ${\rm{diag}}(\bm{v})$ constructs a diagonal matrix using $\bm{v}$ as diagonal elements, while for a matrix $\bm{M}$, ${\rm{diag}}(\bm{M})$ obtains diagonal elements and reshapes them into a corresponding vector.
\section{System Model and Problem Formulation}

\begin{figure}[t]
\centering
\includegraphics[width=0.45\textwidth]{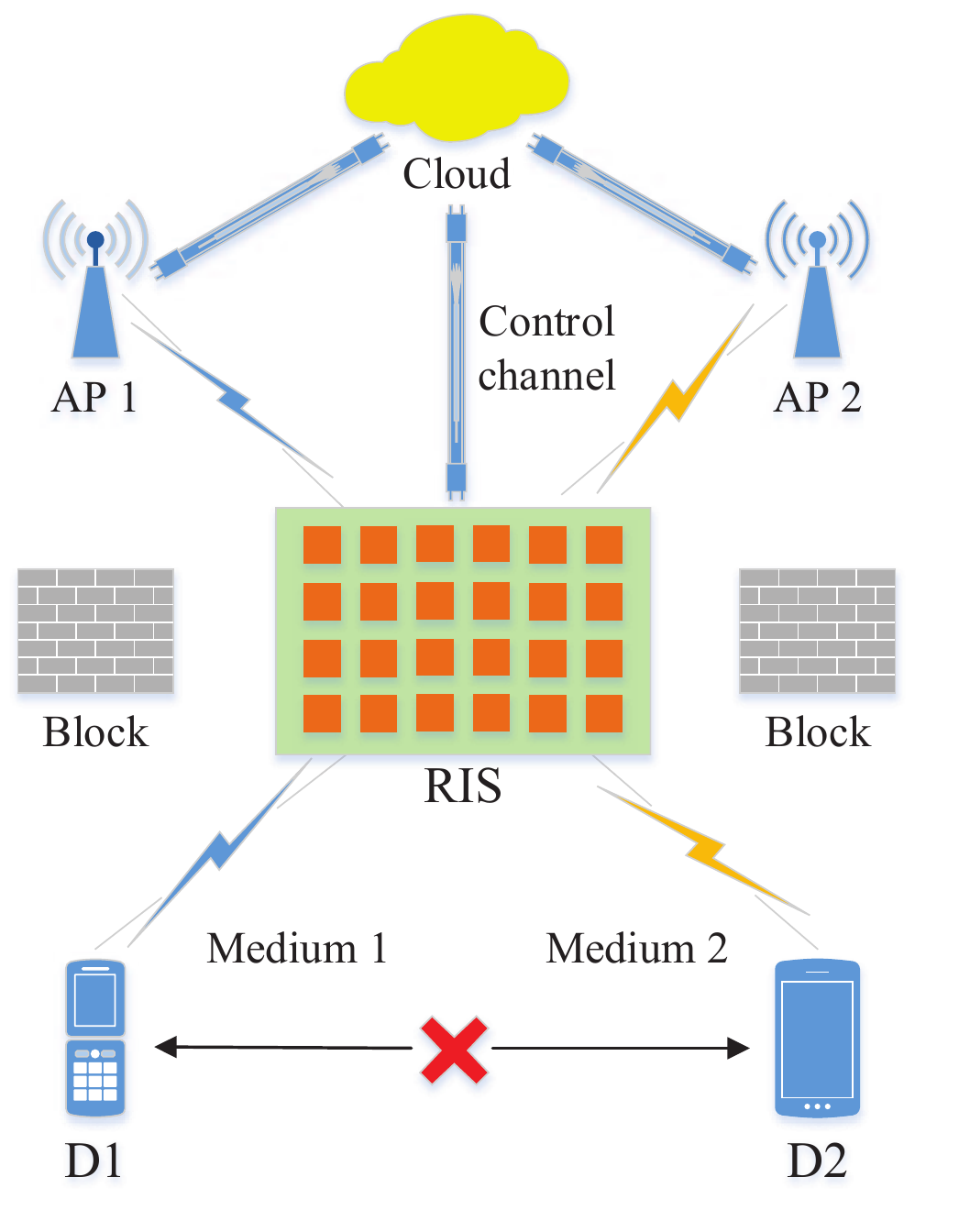}  %w2.5in h3in
\caption{ General structure of RIS-aided cross-media communications.}
\label{Fig1}
\end{figure}

\subsection{System Description}
Assume a cross-media communications scenario, where two devices using heterogenous media, denoted by D1 and D2, ask for information exchange with each other, while each of AP1 and AP2 connected to the cloud, only supports the access to a single medium, e.g., medium 1 and 2, respectively. Due to the lack of multi-media modems, device-to-device communications between D1 and D2 is not available. In addition, it is assumed that there is not a direct link between devices and the APs. According to the RIS features described in \cite{BasarE-2019}, the RIS has full-band response ideally, which induces variable phase-shift responses at different frequency bands. \footnote{Based on the shared-aperture method proposed in \cite{LiT-2018}, the dual-band RIS can be theoretically achieved by integrating two unit elements of different frequency bands in the same metasurface. However, how to implement the smart configuration of the dual-band RIS for cross-media communications deserves further investigation.} In practical implementations, the EM response of RIS over different frequency bands may be different \cite{CaiW-2022}. Without loss of generality, we assume that the differences in magnitude response between media 1 and 2 are generally indicated by different antenna gains. Meanwhile, the differences in phase shift response can be uniformly compensated at the receiver side, since each RE has the same reflection phase offset between media 1 and 2. Under the hypothesis of the RIS working at different media ideally, system framework is designed and theoretical performance is evaluated for the considered RIS-aided cross-media communication system.

In comparison with the conventional relay, the RIS can reduce the hardware cost and power consumption meanwhile the identical structure of REs at different frequency bands brings a unique advantage of scalability to cross-media communications. Therefore, the RIS is coupled with cross-media communications in order to assist the process of end-to-end information exchange in an energy-efficient, low-cost and scalable manner. The general structure of RIS-aided cross-media communications is shown in Fig. \ref{Fig1}. In the uplink transmission, the RIS-aid reflection structure is utilized for providing multipath gains. In the downlink transmission, the RIS-based transmitter is recommended for low-cost implementations of a MIMO transmitter. Since the RIS is passive, it requires the incident carriers for downlink transmission. Although the incident carriers or the downlink signals can also be transmitted from the AP, i.e. AP-based downlink transmission, we tend to let the devices transmit incident carriers to the RIS, since the known carrier signals can reduce the complexity of synchronization at the devices. In addition, when devices are placed in the near field of the RIS, the RIS-based transmitter can provide higher energy efficiency (EE), compared to the AP-based downlink transmission. Based on the above structure, the cloud-management protocol for RIS-aided cross-media communications is proposed. Firstly, D1 and D2 transmit their data to the corresponding APs, with the aid of RIS-aided dual-hop transmission. Thanks to the heterogeneity of propagation media, the uplink signals from different devices can be naturally distinguished and will not generate mutual interferences at the APs. Then, the APs implement the intermediate frequency conversion and upload the signals to the cloud via the optic fiber. After demodulating and decoding, the cloud obtains the digital baseband of D1 and D2 transmit signals. Combined with the additional control information, the baseband signals will be transformed into the control signals, as the input to the RIS phase-shift controller with the aid of the DAC. Thereby, using the RIS as a passive transmitter and transmitting the incident carriers to the RIS, devices can read the required information from the reflected signals. Moreover, by contrast to the medium-divided uplink signals, the downlink signals are distinguished base on TDMA, that is, the required data is stored in the RIS and read during its preallocated slots. Hence, cross-media information exchange can be achieved without the participation of the APs in the downlink, leading to reduced energy consumption.

Let us assume that each device has a transceiver working at SISO mode for full-duplex operation and an additional antenna for carrier transmission to the RIS, while each AP is equipped with $M$ receive antennas. To improve the transmission efficiency, devices adopt the co-time and co-frequency full-duplex mode. Since the APs dispense with any operation in the downlink transmission, the half-duplex mode is adopted at the AP. Although the full-duplex mode imposes self-interferences (SI) between uplink and downlink signals and loop interferences (LI) between transmit and receive antennas, the AP can effectively eliminate SI and LI whereas the hardware-limited devices remain the part of inferences, with the aid of full-duplex interference mitigation techniques \cite{RiihoT-2011,EvereE-2014}.
\subsection{Signal Model}
The channel state information (CSI) of all channels is assumed to be frequency-flat, quasi-static and perfectly known to the APs. \footnote{In order to characterize the theoretical performance bound of the proposed system, we simplify the process of the CSI acquisition and assume that the CSI is perfectly known to the AP ideally. In practical scenarios, perfect channel estimation may be carried out by the embedded low-power sensors in the RIS \cite{RenzoMD-2019}. Another low-complexity solution is to estimate the cascaded (or product) channel among the transmitter, the RIS and the the receiver rather than all of individual channels \cite{HeZ-2020}. The effects of channel estimation errors on system performance are demonstrated in the following section of numerical simulations.} Defining the device set as $D = \{ 1,2\}$, the signal transmitted from $D_i$ is expressed as
\begin{eqnarray}\label{Eq1}
\begin{aligned}
{x_i}(t) = \sqrt {{P_i}} {s_i}(t)\cos 2\pi {f_i}t,i \in D,
\end{aligned}
\end{eqnarray}
where ${P_i}$ is the transmit power, ${s_i}(t)$ denotes the unit-power transmit symbol, ${f_i}$ is the carrier frequency of medium $i$. The equivalent baseband of ${x_i}(t)$ is expressed as $x_i^l(t) = {e^{j{w_{{m_i}}}(t)}}$, where ${w_{{m_i}}} = 2\pi (m - 1)/M,m = 1,2,...,M,$ is selected from M-ary PSK constellation diagram.

Considering the RIS has $N$ reflection elements (REs) with adjustable phases, we denote the phase-shifting vector by ${{\bm{\Phi}} ^T} = [{e^{j{\theta _1}}},{e^{j{\theta _2}}},...,{e^{j{\theta _N}}}]$, consisting of ${\theta _i} = {\phi _i} + w_m^{{\rm{RIS}}}$ where ${\phi _i}$ indicates the channel gain-related phase term and $w_m^{{\rm{RIS}}}$ is the information-dependent phase term. The diagonal matrix form of ${\bm{\Phi}}$ is given by ${\rm{diag}}( {\bm{\Phi}} ) = {e^{jw_m^{{\rm{RIS}}}}} \cdot {\rm{diag}}({\bm{\phi}} )$, where ${\rm{diag}}({\bm{\phi}} )$ is the diagonal matrix with the diagonal element ${e^{j{\phi _i}}},i = 1,2,...,N.$

Then, the discrete-time signal received by the ${\rm{AP}}_i$ is expressed as
\begin{eqnarray}\label{Eq2}
\begin{aligned}
{\bm{y}}_i^{\rm{U}} =& \sqrt {{P_i}} {{\bm{H}}_{i,0}} {\rm{diag}}( {\bm{\Phi}} ) {{\bm{h}}_{i,r}} x_i^l + {{\bm{n}}_0}\\
 =& \sqrt {{P_i}} {{\bm{H}}_{i,0}} {\rm{diag}}( {\bm{\phi}} ) {{\bm{h}}_{i,r}}{e^{jw_m^{{\rm{RIS}}}}} x_i^l + {{\bm{n}}_0},
\end{aligned}
\end{eqnarray}
where ${\pmb{h}_{i,r}} \in {\mathbb{C}^{N \times 1}}, {\bm{H}_{i,0}} \in {\mathbb{C}^{M \times N}}$ represent the baseband equivalent channel from $D_i$ to the RIS and that from the RIS to the ${\rm{A}}{{\rm{P}}_i}$, respectively, $\bm{n}_0 \in {\mathbb{C}^{M \times 1}}$ is the additive white gaussian noise (AWGN) with mean 0 and variance $\sigma _0^2$.

Since the cloud can send the feedback information of $w_m^{{\rm{RIS}}}$ to the APs via the backhaul link, the interferences introduced by the RIS-carried messages can be mitigated by the product of  ${\bm{y}}_i^{\rm{U}}$ and ${e^{ - jw_m^{{\rm{RIS}}}}}$. In addition, the minimum mean square error (MMSE) decoding rule is utilized for signal detection based on the receive beamforming vector denoted by ${\bm{w}}_i^{\rm{U}} \in {\mathbb{C}^{M \times 1}}$. The expression of ${\bm{w}}_i^{\rm{U}}$ and the MSE of decoding signal will be further analyzed in the subsequent section.

Thus, the estimated signal, denoted by $\hat s_i^{\rm{U}}$, is expressed as
\begin{eqnarray}\label{Eq3}
\begin{aligned}
\hat s_i^{\rm{U}} =& {({\bm{w}}_i^{\rm{U}})^H} {e^{ - jw_m^{{\rm{RIS}}}}} {\bm{y}}_i^{\rm{U}}\\
 =& {({\bm{w}}_i^{\rm{U}})^H}(\sqrt {{P_i}} {{\bm{H}}_{i,0}}{\rm{diag}}({\bm{\phi}} ){{\bm{h}}_{i,r}} x_i^l + {e^{ - jw_m^{{\rm{RIS}}}}}{{\bm{n}}_0}).
\end{aligned}
\end{eqnarray}

From (\ref{Eq3}), the received signal-to-interference-plus-noise ratio (SINR) at the ${\rm{AP}}_i$ can be formulated as
\begin{eqnarray}\label{Eq4}
\begin{aligned}
\gamma _i^{\rm{U}} = \frac{{{P_i}{{\left| {{{({\bm{w}}_i^{\rm{U}})}^H}{{\bm{H}}_{i,0}} {\rm{diag}}({\bm{\phi}}){{\bm{h}}_{i,r}}} \right|}^2}}}{{{{\left| {({\bm{w}}_i^{\rm{U}})} \right|}^2}\sigma _0^2}}.
\end{aligned}
\end{eqnarray}

To reduce the power consumption at the AP side and the complexity of synchronization at the device side, the RIS-based transmitter is used for downlink transmission. The baseband signals at the cloud are transmitted together with phase-shifting information, as control signals of the RIS. The devices can read the required information from the reflected signals by transmitting the incident carries to the RIS. Under the assumption of full-duplex operation, the downlink received signals suffer from the self-interference (SI) introduced by the reflected $x_i$ from the RIS and the loop interference (LI) between the local transmit and receive antennas.

Thus, the discrete-time signal received by $D_i$ can be written as
\begin{eqnarray}\label{Eq5}
\begin{aligned}
y_i^{\rm{D}} =& \sqrt {{P_i}} {{\bm{g}}_{i,r}}{\rm{diag}}({\bm{\Phi}} ){{\bm{h}}_{i,r}} + \sqrt {{P_i}} {{\bm{g}}_{i,r}}{\rm{diag}}({\bm{\Phi}} ){{\bm{h}}_{i,r}}x_i^l + \\
&\sqrt {{P_i}} {h_{ii}}x_i^l + {n_i}\\
 =& \sqrt {{P_i}} {{\bm{g}}_{i,r}}{\rm{diag}}({\bm{\phi}} ){{\bm{h}}_{i,r}}x_j^l + \sqrt {{P_i}} {{\bm{g}}_{i,r}}{\rm{diag}}({\bm{\phi}} ){{\bm{h}}_{i,r}}x_j^lx_i^l + \\
 &\sqrt {{P_i}} {h_{ii}}x_i^l + {n_i},
\end{aligned}
\end{eqnarray}
where $ {\bm{g}}_i^r \in {\mathbb{C}^{1 \times N}}$ is the baseband equivalent channel from the RIS to $D_i$, the first term is the reflected signals from the incident carriers to the RIS, the second term is the SI from the reflected ones of  $x_i^l$, the third term indicates the LI and $h_{ii}$ is the loop channel between the transmit and receive antenna.

Since the local signal $x_i^l$, the cascade channel ${{\bm{g}}_{i,r}} {\bm{\Phi}} {{\bm{h}}_{i,r}}$ and the loop channel $h_{ii}$ are known to the receiver of $D_i$, interference suppression can be used for reducing SI and LI \cite{ZhongC-2014}. The recovered signal after interference suppression, denoted by $\hat y_i^{\rm{D}}$, is expressed as \cite{PengZ-2021}
\begin{eqnarray}\label{Eq6}
\begin{aligned}
\hat y_i^{\rm{D}} =& \sqrt {{P_i}} {{\bm{g}}_{i,r}} {\rm{diag}} ({\bm{\Phi}}) {{\bm{h}}_{i,r}} +\sqrt {{\rho _{\rm{SI}}}{P_i}} {{\bm{g}}_{i,r}} {\bm{\Phi}} {{\bm{h}}_{i,r}}x_i^l + \\
&\sqrt {{\rho _{\rm{LI}}}{P_i}} {h_{ii}}x_i^l + {n_i}\\
 =& \sqrt {{P_i}} {{\bm{g}}_{i,r}}{\rm{diag}}({\bm{\phi}} ){{\bm{h}}_{i,r}}x_j^l + \sqrt {{\rho _{\rm{SI}}}{P_i}} {{\bm{g}}_{i,r}} {\bm{\Phi}} {{\bm{h}}_{i,r}}x_i^l + {{\tilde n}_i},
\end{aligned}
\end{eqnarray}
where the second term indicates the residual part of SI and ${\rho _{{\rm{SI}}}}$ is the SI coefficient, the third term is the residual part of LI and ${\rho _{{\rm{LI}}}}$ is the LI coefficient, ${\tilde n_i} = \sqrt {{\rho _{{\rm{LI}}}}{P_i}} {h_{ii}}x_i^l + {n_i}$ denotes the LI-plus-noise item with average power $\tilde \sigma _i^2 = {\rho _{{\rm{LI}}}}{P_i}{\left| {{h_{ii}}} \right|^2} + \sigma _i^2$.

Thereby, the downlink received SINR at $D_i$ can be expressed as
\begin{eqnarray}\label{Eq7}
\gamma _i^D = \frac{{{P_i}{{\left| {{\bm{g}_{i,r}} {\bm{\phi}} {{\pmb{h}_{i,r}}}} \right|}^2}}}{{{\rho _{\rm{SI}}}{P_i}{{\left| {{\bm{g}_{i,r}} {\bm{\phi}} {{\pmb{h}_{i,r}}}} \right|}^2} + \tilde \sigma _i^2}}.
\end{eqnarray}

Finally, the Shannon capacity of the uplink and the downlink channel is denoted by $C = [C_1^{\rm{U}},C_2^{\rm{U}},$ $C_1^{\rm{D}},C_2^{\rm{D}}]$, respectively, given by
\begin{subequations}\label{Eq8}
\begin{align}
C_i^{\rm{U}} =& {B_i}\log (1 + \gamma _i^{\rm{U}}),\\
C_i^{\rm{D}} =& {B_i}\log (1 + \gamma _i^{\rm{D}}),
\end{align}
\end{subequations}
where $B_i$ represents the channel bandwidth. Due to the differences of channel bandwidth between heterogenous media, $B_i$ should not be normalized.

\subsection{Problem Formulation}
 In this paper, we formulate the cross-media information exchange model as a optimization problem of maximizing the end-to-end throughput by jointly optimizing the RIS phase, transmit rate and time allocation in the cases with and without delay constraints.

\subsubsection{Optimization Model without Delay Constraints}
\begin{figure}[t]
\centering
\includegraphics[width=0.45\textwidth]{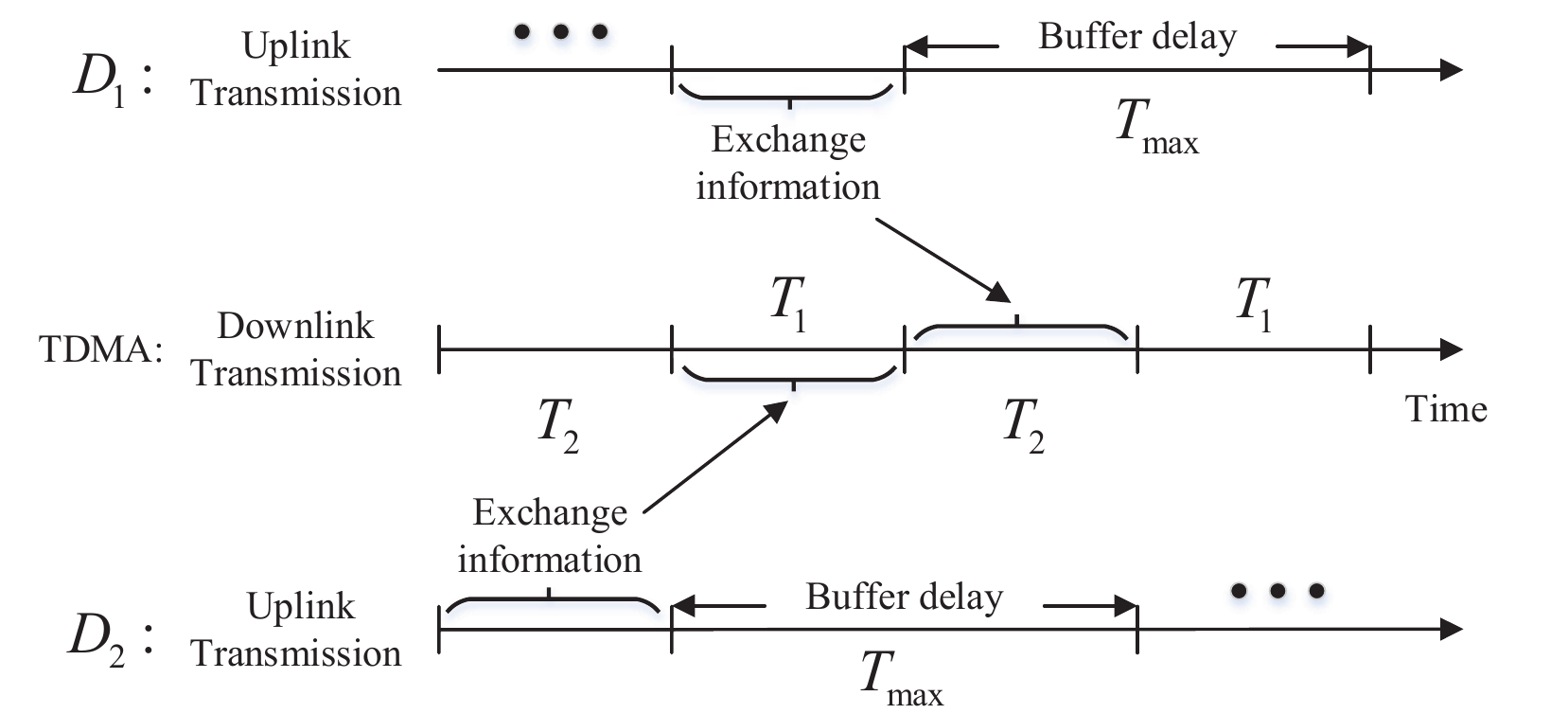}
\caption{ Frame structure of the cloud-management transmission protocol.}
\label{Fig2}
\end{figure}

Let us define the frame length as ${T_{\max }}$  and the vector of time allocation as $T = [{T_1},{T_2}]$ , where $T_i$ is the preallocated slot length for the downlink transmission of $D_i$. Assuming the vector of transmit rate as $R = [R_1^{\rm{U}},R_2^{\rm{U}},R_1^{\rm{D}},R_2^{\rm{D}}]$, the end-to-end throughput for a full-duplex relaying system can be expressed as \cite{AtapaS-2013}
\begin{subequations}\label{Eq9}
\begin{align}
{S_{1 \to 2}} =& \min \{ R_1^{\rm{U}},R_2^{\rm{D}}\}  \cdot {T_2},\\
{S_{2 \to 1}} =& \min \{ R_2^{\rm{U}},R_1^{\rm{D}}\}  \cdot {T_1}.
\end{align}
\end{subequations}

Considering the fairness of information exchange, the optimization objective is set to maximize the minimal value of the end-to-end throughput. The RIS phase, transmit rate and time allocation are jointly optimized for dealing with the mentioned cross-media information exchange issue, which can finally be formulated as
\begin{subequations}\label{Eq10}
\begin{align}
\textbf{(P1)}~~&\mathop {\max \min }\limits_{R,T,{\bm{\phi}} } {\rm{ \{ }}{S_{1 \to 2}},{S_{2 \to 1}}\} \\
{\rm{s}}{\rm{.t}}{\rm{. }}~
&R \le C({\bm{\phi}} ),\\
&{T_1} + {T_2} = {T_{\max }},\\
&\left| {{{\bm{\phi}} _n}} \right| = 1,\forall n = 1,...,N,
\end{align}
\end{subequations}
where the constraint (b) and (c) give a boundary condition of the transmit rate and time allocation, based on the channel capacity and the frame length, respectively. The constraint (d) indicates the limitation of the normalized energy for each RIS reflection unit.

For notational simplicity, the frame-normalized operation is carried out and the vector of time allocation is redefined as ${T_i} \buildrel \Delta \over = {T_i}/{T_{\max }}$. In addition, due to the relationship between the transmit rate and channel capacity shown in the objective and the constraint (b) in Eq. (\ref{Eq10}), it can be naturally observed that \textbf{(P1)} can be rewritten as
\begin{subequations}\label{Eq11}
\begin{align}
\textbf{(P1.1)}~~
&\mathop {\max \min }\limits_{T,{\bm{\phi}} } {\rm{ \{ }}C_1^{\rm{U}}{T_2},C_2^{\rm{U}}{T_1},C_1^{\rm{D}}{T_1},C_2^{\rm{D}}{T_2}\} \\
{\rm{s}}{\rm{.t}}{\rm{.}}~
&R_1^{\rm{U}} = R_2^{\rm{D}} = \min \{ C_1^{\rm{U}},C_2^{\rm{D}}\} ,\\
&R_2^{\rm{U}} = R_1^{\rm{D}} = \min \{ C_2^{\rm{U}},C_1^{\rm{D}}\} ,\\
&{T_1} + {T_2} = 1,\\
&\left| {{{\bm{\phi}} _n}} \right| = 1,\forall n = 1,...,N.
\end{align}
\end{subequations}

\subsubsection{Optimization Model with Delay Constraints}
Limited to the cache size and low-latency demand under the delay-sensitive scenarios, the process of information exchange should be finished before the tolerable delay. Without loss of generality, we assume that the tolerable delay of both devices is identical and equal to ${T_{\max }}$. As shown in Fig. \ref{Fig2}, the uplink transmission of $\rm{D}1$ and $\rm{D}2$ is implemented simultaneously while the downlink transmission is based on TDMA. During the pre-allocated slot $T_1$, $\rm{D}1$ can receive the exchanged information transmitted from $\rm{D}2$ during the last $T_1$ slot. Similarly, during the pre-allocated slot $T_2$, $\rm{D}2$ can receive the exchanged information transmitted from $\rm{D}1$ during the last $T_1$ slot. Thereby, the maximum buffer delay can be controlled in the range of the frame length $T_{\rm{max}}$. To satisfy such delay constraints, the relationship between the transmit rate and time allocation should arrive at
\begin{subequations}\label{Eq12}
\begin{align}
R_1^{\rm{D}}{T_1} & \ge R_2^{\rm{U}}{T_2},\\
R_2^{\rm{D}}{T_2} & \ge R_1^{\rm{U}}{T_1}.
\end{align}
\end{subequations}
where we observe that the difference between the uplink rate of one device and the downlink rate of the other one has important effects on the delay constraints. Hence, the rate discrepancy between heterogeneous media prompts us to adopt rate adaption techniques to meet the delay constraints.

 After adding the delay constraints (\ref{Eq12}) into Problem \textbf{(P1)} and implementing the frame-normalized operation, the optimization problem with delay constraints is modeled as
\begin{subequations}\label{Eq13}
\begin{align}
\textbf{(P2)}~~
&\mathop {\max \min }\limits_{R,T,{\bm{\phi}} } {\rm{ \{ }}{S_{1 \to 2}},{S_{2 \to 1}}\} \\
{\rm{s}}{\rm{.t}}{\rm{. }}~
&R_i^{\rm{U}}{T_i} \le R_j^{\rm{D}}{T_j},i \ne j\\
&R \le C({\bm{\phi}} ),\\
&{T_1} + {T_2} = 1,\\
&\left| {{{\bm{\phi}} _n}} \right| = 1,\forall n = 1,...,N.
\end{align}
\end{subequations}

\section{Proposed Algorithm}

In the mathematical model of problem $\textbf{(P1)}$ and $\textbf{(P2)}$, the RIS phase, transmit rate and time allocation are jointly optimized for the max-min objective function of the end-to-end throughput. The key challenges of addressing the above problems focus on the heterogenous characteristics of different media, the non-convexity of the objective function and the coupling of different optimization variables. Due to the mentioned issues of cross-media communications, the existing method cannot be directly extended to our model, which requires further developments such as the non-convex optimization of the objective function, the operation of decoupling optimization variables and the tradeoff of transmit rate between heterogeneous media, as shown in the proposed algorithm in the following.
\subsection{Problem Transformation}
Since $C({\bm{\phi}} )$ is a non-convex function of the variable ${\bm{\phi}}$, we firstly deal with such non-convexity in order to solve \textbf{(P1)} and \textbf{(P2)} further. Under the assumption of quasi-static channels with perfect-known CSI and ${\bm{\phi}} $, the minimum mean square error (MMSE) equalizer is recommended for maximizing the received SINR. The relationship between the channel capacity and MMSE equalization is utilized for substituting $C({\bm{\phi}} )$ with an equivalent convex function form. The related details are shown as follows.

According to (\ref{Eq3}) and (\ref{Eq6}), the cascade channel of the uplink and downlink transmission can be respectively written as
\begin{subequations}\label{Eq14}
\begin{align}
\bar {\bm{h}}_i^{\rm{U}} \buildrel \Delta \over =& \sqrt {{P_i}} {\bm{H}_{i,0}}{\rm{diag}}({\bm{\phi}} ){{\pmb{h}_{i,r}}},\\
\bar h_i^{\rm{D}} \buildrel \Delta \over =& \sqrt {{P_i}} {\bm{g}_{i,r}}{\rm{diag}}({\bm{\phi}} ){{\pmb{h}_{i,r}}}.
\end{align}
\end{subequations}

Defining the receive beamforming vector of downlink transmission as $w_i^{\rm{D}}$, the estimated signal is given by $\hat s_i^{\rm{D}} = {(w_i^{\rm{D}})^H} y_i^{\rm{D}}$. Based on the MMSE criteria, the receive beamforming vector can be computed by
\begin{subequations}\label{Eq15}
\begin{align}
\bm{w}_i^{\rm{U}} =& \bar {\bm{h}}_i^{\rm{U}}{(\bar {\bm{h}}_i^{\rm{U}}{(\bar {\bm{h}}_i^{\rm{U}})^H} + \sigma _0^2{\bm{I}_M})^{ - 1}},\\
w_i^{\rm{D}} =& \bar h_i^{\rm{D}}{(\bar h_i^{\rm{D}}{(\bar h_i^{\rm{D}})^H} + {\rho _{\rm{SI}}}{P_i}{\left| {{\bm{g}_{i,r}}{\rm{diag}}({\bm{\phi}} ){{\pmb{h}_{i,r}}}} \right|^2} + \tilde \sigma _i^2)^{ - 1}}.
\end{align}
\end{subequations}

Thereby, the MSE of decoding symbols can be derived by
\begin{subequations}\label{Eq16}
\begin{align}
\begin{split}
e_i^{\rm{U}}({\bm{\phi}}, \bm{w}_i^{\rm{U}}) =& E[(\hat s_i^{\rm{U}} - s_i^{\rm{U}}){(\hat s_i^{\rm{U}} - s_i^{\rm{U}})^H}]\\
=& ({(\bm{w}_i^{\rm{U}})^H}\bar {\bm{h}}_i^{\rm{U}} - 1){({(\bm{w}_i^{\rm{U}})^H}\bar {\bm{h}}_i^{\rm{U}} - 1)^H} + \\
&{(\bm{w}_i^{\rm{U}})^H}\bm{w}_i^{\rm{U}}\sigma _0^2,\\
\end{split}
\end{align}
\begin{align}
\begin{split}
e_i^{\rm{D}}({\bm{\phi}}, w_i^{\rm{D}}) =& E[(\hat s_i^{\rm{D}} - s_i^{\rm{D}}){(\hat s_i^{\rm{D}} - s_i^{\rm{D}})^H}]\\
 =& ({(w_i^{\rm{D}})^H}\bar h_i^{\rm{D}} - 1){({(w_i^{\rm{D}})^H}\bar h_i^{\rm{D}} - 1)^H} + \\
 &{(w_i^{\rm{D}})^H}w_i^{\rm{D}}\tilde \sigma _i^2+ {\rho _{\rm{SI}}}{P_i}{(w_i^{\rm{D}})^H}\bar h_i^{\rm{D}}{(\bar h_i^{\rm{D}})^H}w_i^{\rm{D}}.
\end{split}
\end{align}
\end{subequations}

In addition, the relationship between the receive rate and the MSE based on MMSE equalization is utilized for converting $C({\bm{\phi}} )$ into the equivalent form as follows \cite{PanC-2020}:
\begin{subequations}\label{Eq17}
\begin{align}
\bar C_i^{\rm{U}}({\bm{\phi}}, \bm{w}_i^{\rm{U}}, \mu _i^{\rm{U}}) =& \log \left| {\mu _i^{\rm{U}}} \right| - \mu _i^{\rm{U}} e_i^{\rm{U}} + 1,\\
\bar C_i^{\rm{D}}({\bm{\phi}}, w_i^{\rm{D}}, \mu _i^{\rm{D}}) =& \log \left| {\mu _i^{\rm{D}}} \right| - \mu _i^{\rm{D}} e_i^{\rm{D}} + 1,
\end{align}
\end{subequations}
where $\mu  = \{ \mu _i^{\rm{U}},\mu _i^{\rm{D}}\}$ are the introduced auxiliary variables. The optimal $\mu$ for given ${\bm{\phi}}$ can be obtained by the first-order derivation, given by
\begin{subequations}\label{Eq18}
\begin{align}
\mu _i^{\rm{U}} =& {(e_i^{\rm{U}})^{ - 1}},\\
\mu _i^{\rm{D}} =& {(e_i^{\rm{D}})^{ - 1}}.
\end{align}
\end{subequations}

For any given ${\bm{\phi}}$, $w = \{ \bm{w}_i^{\rm{U}}, w_i^{\rm{D}}\}$  or $\mu $,  $\bar C({\bm{\phi}}, w ,\mu )$ is a convex function of other variables.
To deal with the non-convexity introduced by the unit-modulus constraint, we replace the constraint with the relaxed one $\left| {{{\bm{\phi}} _n}} \right| \le 1,n = 1,...,N$ in \textbf{(P1.2)} and \textbf{(P2.1)}. The final feasible phase vector $\hat {\bm{\phi}} ^{\rm{opt}}$ can be approximate to $\deg ({{\bm{\phi}} ^{{\rm{opt}}}})$, where ${{\bm{\phi}} ^{{\rm{opt}}}}$ is the optimal phase vector under relaxed constraints. To mitigate the effect of taking the approximate $\hat {\bm{\phi}} ^{\rm{opt}} $ on the algorithm convergence, we consider
\begin{eqnarray}\label{Eq32}
\hat {\bm{\phi}} ^{\rm{opt}}(t) =\left\{
\begin{aligned}
&{\deg ({{\bm{\phi}} ^{{\rm{opt}}}}(t)),~{\rm{ if }}~F(\deg ({{\bm{\phi}} ^{{\rm{opt}}}}(t))) \ge F(t - 1),}\\
&{{{\bm{\phi}} ^{{\rm{opt}}}}(t),~{\rm{ if}}~F(\deg ({{\bm{\phi}} ^{{\rm{opt}}}}(t))) < F(t - 1).}
\end{aligned}
 \right.
\end{eqnarray}
where $t$ indicates the iteration number and $F$ represents the objective function of \textbf{(P1.b)}.

Finally, the corresponding relaxed problems of \textbf{(P1)} and \textbf{(P2)} in the case of non-unimodular phase vector can be reformulated as follows:
\begin{subequations}\label{Eq19}
\begin{align}
\textbf{(P1.2)}~~
&\mathop {\max \min }\limits_{\begin{array}{*{20}{c}}
{w,\mu }\\
{T,{\bm{\phi}} }
\end{array}} {\rm{ \{ }}\bar C_1^{\rm{U}}{T_2},\bar C_2^{\rm{U}}{T_1},\bar C_1^{\rm{D}}{T_1},\bar C_2^{\rm{D}}{T_2}\} \\
{\rm{s}}{\rm{.t}}{\rm{. }}~
&R_1^{\rm{U}} = R_2^{\rm{D}} = \min \{ C_1^{\rm{U}},C_2^{\rm{D}}\} ,\\
&R_2^{\rm{U}} = R_1^{\rm{D}} = \min \{ C_2^{\rm{U}},C_1^{\rm{D}}\} ,\\
&{T_1} + {T_2} = 1,\\
&\left| {{{\bm{\phi}} _n}} \right| \le 1,\forall n = 1,...,N.
\end{align}
\end{subequations}

and

\begin{subequations}\label{Eq20}
\begin{align}
\textbf{(P2.1)}~~
&\mathop {\max \min }\limits_{\begin{array}{*{20}{c}}
{w,\mu }\\
{R,T,{\bm{\phi}} }
\end{array}} {\rm{ \{ R}}_1^{\rm{U}}{T_2},R_2^{\rm{U}}{T_1},R_1^{\rm{D}}{T_1},R_2^{\rm{D}}{T_2}\} \\
{\rm{s}}{\rm{.t}}{\rm{. }}~
&R_i^{\rm{U}}{T_i} \le R_j^{\rm{D}}{T_j},i \ne j\\
&R \le \bar C({\bm{\phi}} ,w,\mu ),\\
&{T_1} + {T_2} = 1,\\
&\left| {{{\bm{\phi}} _n}} \right| \le 1,\forall n = 1,...,N.
\end{align}
\end{subequations}

\begin{figure}[t]
\centering
\subfigure[Algorithm framework of module-divided iterative optimization for \textbf{(P1)}]{
\centering
\includegraphics[width=0.35\textwidth]{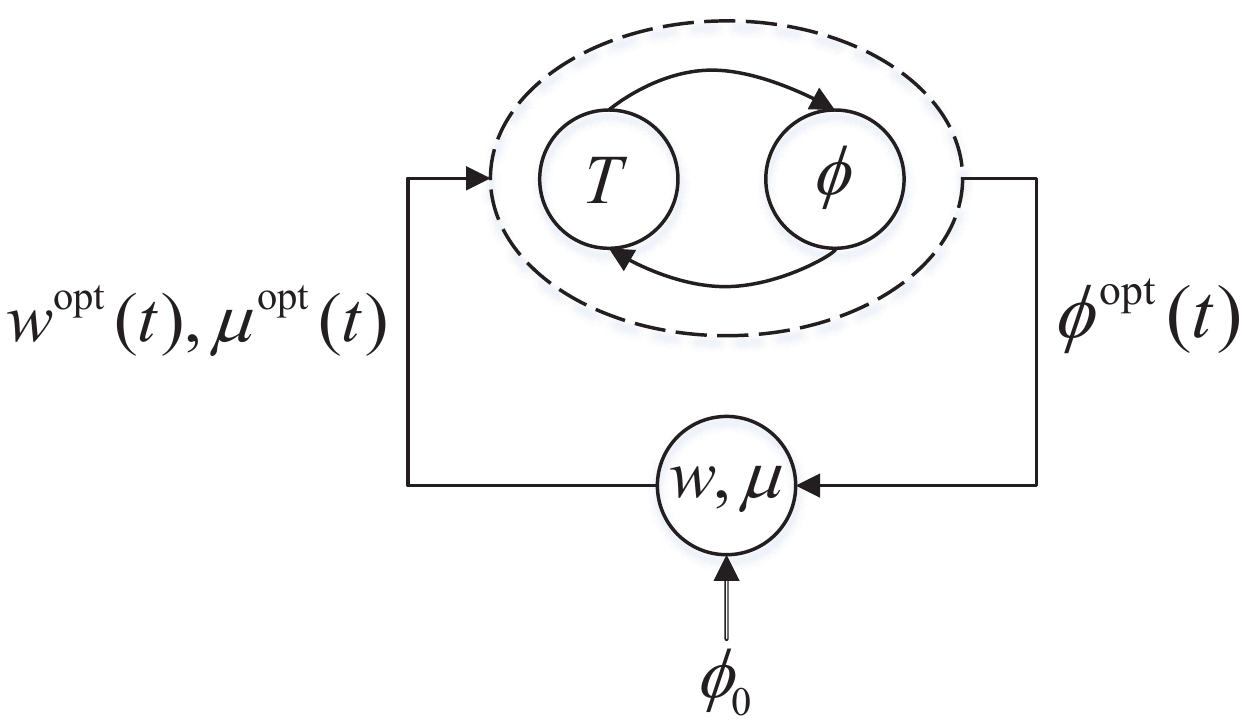} }  %0.4
\subfigure[Algorithm framework of module-divided iterative optimization for \textbf{(P2)}]{
\centering
\includegraphics[width=0.35\textwidth]{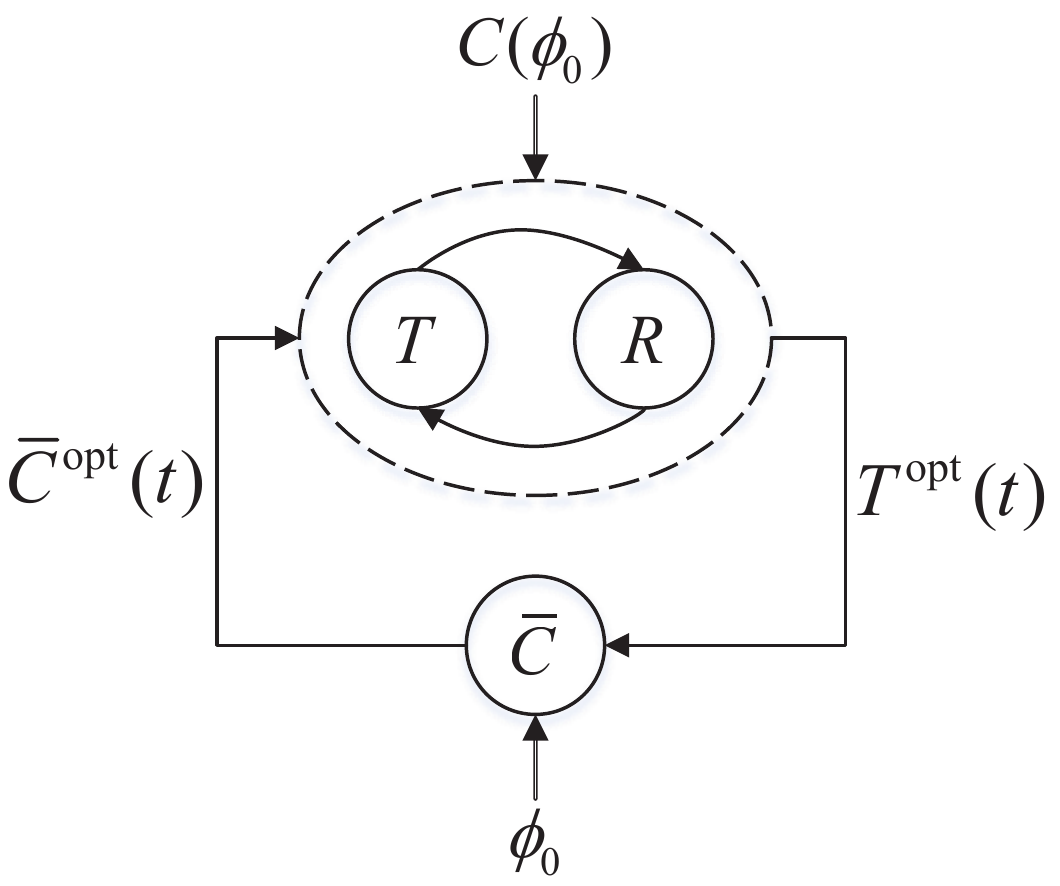} } %0.32
\caption{General framework of the proposed algorithm.}
\label{Fig3}
\end{figure}
\subsection{Proposed Algorithm for \textbf{(P1)}}
As shown in Fig. \ref{Fig3}(a), the process of iterative optimization for \textbf{(P1.2)} is updated between two independent modules, e.g., the joint optimization of $T, {\bm{\phi}}$ including two sub-modules and the acquisition of ${w^{{\rm{opt}}}}, {\mu ^{{\rm{opt}}}}$. Moreover, since the optimal $w, \mu $ for given ${{\bm{\phi}} ^{{\rm{opt}}}}$ can be directly obtained by Eq. (\ref{Eq15}) and Eq. (\ref{Eq18}), the computational complexity mainly includes the joint optimization of $T, {\bm{\phi}} $ for given ${w^{{\rm{opt}}}}, {\mu ^{{\rm{opt}}}}$. The process of each sub-module optimization is respectively shown as follows.

\subsubsection{Optimizing Time Allocation $T$}
For given $w, \mu $, the joint optimization problem of $T, {\bm{\phi}} $ can be modeled as
\begin{subequations}\label{Eq21}
\begin{align}
\textbf{(P1.a)}~~
&\mathop {\max }\limits_T \underbrace {\mathop {\max }\limits_{\bm{\phi}}  \min \{ \bar C_1^{\rm{U}}{T_2},\bar C_2^{\rm{U}}{T_1},\bar C_1^{\rm{D}}{T_1},\bar C_2^{\rm{D}}{T_2}\} }_{{\rm{Sub - problem~of~optimizing~}}{\bm{\phi}} }\\
{\rm{s}}{\rm{.t}}{\rm{. }}~
&R_1^{\rm{U}} = R_2^{\rm{D}} = \min \{ C_1^{\rm{U}},C_2^{\rm{D}}\} ,\\
&R_2^{\rm{U}} = R_1^{\rm{D}} = \min \{ C_2^{\rm{U}},C_1^{\rm{D}}\} ,\\
&{T_1} + {T_2} = 1,\\
&\left| {{{\bm{\phi}} _n}} \right| \le 1,\forall n = 1,...,N.
\end{align}
\end{subequations}

Removing the outer optimization of $T$, the inner layer of the above model is related to the sub-module of optimizing ${\bm{\phi}}$. Assuming that the minimum step of time allocation is $\Delta t = 1/{k_{\max }}$, we can utilize one-dimensional search to find ${T^{{\rm{opt}}}}$ and the searching path is given by
\begin{subequations}\label{Eq22}
\begin{align}
{T_1}(k) =& k \cdot \Delta t,\\
{T_2}(k) =& 1 - {T_1}(k),k = 1,...,k_{\max}.
\end{align}
\end{subequations}

\subsubsection{Optimizing the RIS Phase ${\bm{\phi}}$}
For given $w_i^{\rm{U}}$, substituting Eq. (\ref{Eq14}) into Eq. (\ref{Eq16}) and expanding the expression, the MSE of the uplink received signal can be expressed as
\begin{eqnarray}\label{Eq23}
\begin{aligned}
e_i^U({\bm{\phi}} ) =& {P_i}{(\bm{w}_i^{\rm{U}})^H}{\bm{H}_{i,0}} {\bm{\phi}} {{\bm{h}_{i,r}}}{\bm{h}_{i,r}^H}{{ {\bm{\phi}} }^H}\bm{H}_{i,0}^H\bm{w}_i^{\rm{U}} -\\
 &2{\mathop{\rm Re}\nolimits}\{\sqrt {{P_i}} {(\bm{w}_i^{\rm{U}})^H}{\bm{H}_{i,0}} {\bm{\phi}} {{\bm{h}_{i,r}}}\}  + \\
 &{(\bm{w}_i^{\rm{U}})^H}\bm{w}_i^{\rm{U}}\sigma _0^2 + 1\\
 =& {\rm{Tr}}({P_i}{{ {\bm{\phi}} }^H}\bm{H}_{i,0}^H\bm{w}_i^{\rm{U}}{(\bm{w}_i^{\rm{U}})^H}{\bm{H}_{i,0}} {\bm{\phi}} {{\pmb{h}_{i,r}}}{\pmb{h}_{i,r}^H}) - \\
  &2{\mathop{\rm Re}\nolimits}\{{\rm{Tr}}(\sqrt {{P_i}} {{\bm{h}_{i,r}}}{(\bm{w}_i^{\rm{U}})^H}{\bm{H}_{i,0}} {\bm{\phi}} )\}  \\
  &+ {(\bm{w}_i^{\rm{U}})^H}\bm{w}_i^{\rm{U}}\sigma _0^2 + 1.
\end{aligned}
\end{eqnarray}

For notational simplicity, we have
\begin{subequations}\label{Eq24}
\begin{align}
{{\bar {\bm{H}}}_{i,0}} \buildrel \Delta \over =& \bm{H}_{i,0}^H\bm{w}_i^{\rm{U}}{(\bm{w}_i^{\rm{U}})^H}{\bm{H}_{i,0}},\\
{{\bar {\bm{H}}}_{i,r}} \buildrel \Delta \over =& {P_i}{{\bm{h}_{i,r}}}{\bm{h}_{i,r}^H},\\
{{\bar {\bm{H}}}_{i,r,0}} \buildrel \Delta \over =& {\rm{diag}}(\sqrt {{P_i}} {{\bm{h}_{i,r}}}{(\bm{w}_i^{\rm{U}})^H}{\bm{H}_{i,0}}).
\end{align}
\end{subequations}

Then, $e_i^{\rm{U}}({\bm{\phi}} )$ can be reformulated as
\begin{eqnarray}\label{Eq25}
\begin{aligned}
e_i^{\rm{U}}({\bm{\phi}} ) =& {{\bm{\phi}} ^H}({\bar {\bm{H}}_{i,0}} \odot \bar {\bm{H}}_{i,r}^T){\bm{\phi}}  - 2{\mathop{\rm Re}\nolimits} \{ \bar {\bm{H}}_{i,r,0}^T{\bm{\phi}} \}  + \\
&{(\bm{w}_i^{\rm{U}})^H}\bm{w}_i^{\rm{U}}\sigma _0^2 + 1.
\end{aligned}
\end{eqnarray}

For given $w_i^{\rm{D}}$, the MSE of the downlink received signal can be expressed as
\begin{eqnarray}\label{Eq26}
\begin{aligned}
e_i^{\rm{D}}({\bm{\phi}} ) =& {P_i}{(w_i^{\rm{D}})^*}{\bm{g}_{i,r}} {\bm{\phi}} {{\pmb{h}_{i,r}}}{\pmb{h}_{i,r}^H}{{ {\bm{\phi}} }^H}\bm{g}_{i,r}^H w_i^{\rm{D}} - \\
 &2{\mathop{\rm Re}\nolimits} \{\sqrt {{P_i}}{(w_i^{\rm{D}})^*}{\bm{g}_{i,r}}{\bm{\phi}} {{\pmb{h}_{i,r}}}\}  + \\
 &{(w_i^{\rm{D}})^*}w_i^{\rm{D}}\tilde \sigma _i^2 + 1\\
 =& {\rm{Tr}}({P_i}{{{\bm{\phi}} }^H}\bm{g}_{i,r}^Hw_i^{\rm{D}}{(w_i^{\rm{D}})^*}{\bm{g}_{i,r}} {\bm{\phi}} {{\pmb{h}_{i,r}}}{\pmb{h}_{i,r}}^H) - \\
  &2{\mathop{\rm Re}\nolimits} \{ {\rm{Tr}}(\sqrt {{P_i}} {{\pmb{h}_{i,r}}}{(w_i^{\rm{D}})^*}{\bm{g}_{i,r}} {\bm{\phi}} )\}  + \\
  &{(w_i^{\rm{D}})^*}w_i^{\rm{D}}\tilde \sigma _i^2 + 1.
\end{aligned}
\end{eqnarray}
Defining
\begin{subequations}\label{Eq27}
\begin{align}
{{\bar {\bm{G}}}_{i,r}} =& \bm{g}_{i,r}^Hw_i^D{(w_i^{\rm{D}})^*}{\bm{g}_{i,r}},\\
{{\bar {\bm{G}}}_{i,r,0}} =& {\rm{diag}}(\sqrt {{P_i}} {{\pmb{h}_{i,r}}}{(w_i^{\rm{D}})^*}{\bm{g}_{i,r}}),
\end{align}
\end{subequations}
$e_i^{\rm{D}}({\bm{\phi}} )$ can be reformulated as
\begin{eqnarray}\label{Eq28}
\begin{aligned}
e_i^{\rm{D}}({\bm{\phi}} ) =& {{\bm{\phi}} ^H}({\bar {\bm{G}}_{i,r}} \odot \bar {\bm{H}}_{i,r}^T){\bm{\phi}}  - 2{\mathop{\rm Re}\nolimits} \{ \bar {\bm{G}}_{i,r,0}^T{\bm{\phi}} \}  + \\
&{(w_i^{\rm{D}})^*}w_i^{\rm{D}}\tilde \sigma _i^2 + 1.
\end{aligned}
\end{eqnarray}

Moreover, for given $\mu _i^{\rm{U}}, {T_j}$, substituting Eq. (\ref{Eq25}) into Eq.(\ref{Eq17}), we have $f_i^{\rm{U}}({\bm{\phi}} ) = \bar C_i^{\rm{U}}{T_j}$ as follows
\begin{eqnarray}\label{Eq29}
\begin{aligned}
f_i^{\rm{U}}({\bm{\phi}} ) = 2{\mathop{\rm Re}\nolimits} \{ {\bm{A}}_i^{\rm{U}}{\bm{\phi}} \}  - {{\bm{\phi}} ^H} {\bm{B}}_i^{\rm{U}}{\bm{\phi}}  + {\bm{C}}_i^{\rm{U}},
\end{aligned}
\end{eqnarray}
where

\begin{eqnarray*}\label{Eq29-1}
\begin{aligned}
{\bm{A}}_i^{\rm{U}} =& {T_j}\mu _i^{\rm{U}}\bar {\bm{H}}_{i,r,0}^T,\\
{\bm{B}}_i^{\rm{U}} =& {T_j}\mu _i^{\rm{U}}{{\bar {\bm{H}}}_{i,0}} \odot \bar {\bm{H}_{i,r}^T},\\
{\bm{C}}_i^{\rm{U}} =& {T_j}(\log \left| {\mu _i^{\rm{U}}} \right| + 1) - {T_j}\mu _i^{\rm{U}}({(\bm{w}_i^{\rm{U}})^H}\bm{w}_i^{\rm{U}}\sigma _0^2 + 1).
\end{aligned}
\end{eqnarray*}

Similarly, substituting Eq. (\ref{Eq28}) into Eq. (\ref{Eq17}), we arrive at $f_i^{\rm{D}}({\bm{\phi}} ) = \bar C_i^{\rm{D}}{T_i}$ as follows
\begin{eqnarray}\label{Eq30}
\begin{aligned}
f_i^{\rm{D}}({\bm{\phi}} ) = 2{\mathop{\rm Re}\nolimits} \{ {\bm{A}}_i^{\rm{D}}{\bm{\phi}} \}  - {{\bm{\phi}} ^H} {\bm{B}}_i^{\rm{D}}{\bm{\phi}}  + {\bm{C}}_i^{\rm{D}},
\end{aligned}
\end{eqnarray}
where

\begin{eqnarray*}\label{Eq30-1}
\begin{aligned}
{\bm{A}}_i^{\rm{D}} =& {T_i}\mu _i^{\rm{D}}\bar {\bm{G}}_{i,r,0}^T,\\
{\bm{B}}_i^{\rm{D}} =& {T_i}\mu _i^{\rm{D}}\bar {\bm{G}}_{i,0}} \odot {\bar {\bm{H}}_{i,r}^T,\\
{\bm{C}}_i^{\rm{D}} =& {T_i}(\log \left| {\mu _i^{\rm{D}}} \right| + 1) - {T_i}\mu _i^{\rm{D}}({(w_i^{\rm{D}})^*}w_i^{\rm{D}}\tilde \sigma _i^2 + 1).
\end{aligned}
\end{eqnarray*}

Finally, the sub-module of optimizing ${\bm{\phi}}$ can be modeled as
\begin{subequations}\label{Eq31}
\begin{align}
\textbf{(P1.b)}~~
&\mathop {\max }\limits_{\bm{\phi}}  {\rm{min }}\{ f_1^{\rm{U}},f_2^{\rm{U}},f_1^{\rm{D}},f_2^{\rm{D}}\} \\
{\rm{s}}{\rm{.t}}{\rm{. }}~
&\left| {{{\bm{\phi}} _n}} \right| \le 1, \forall n = 1,...,N..
\end{align}
\end{subequations}

Problem \textbf{(P1.b)} can be directly solved via the convex optimization toolbox such as \emph{fminimax} or adopt the MM algorithm proposed in \cite{ZhouG-2020} to obtain the closed-form solution to the surrogate problem of \textbf{(P1.b)}, according to the smooth approximation (\cite{ZhouG-2020}, Theorem 2) and the surrogate function (\cite{ZhouG-2020}, Theorem 4).

The details of the proposed algorithm for \textbf{(P1)} are shown in Alg. \ref{Alg1}.

\begin{algorithm}[t]
\caption{ The proposed iterative optimization algorithm for \textbf{(P1)}}
\label{Alg1}
\begin{algorithmic}
\REQUIRE ~~\\
Initial parameters ${\bm{\phi}}  = {{\bm{\phi}} _0}$
\ENSURE  ~~\\
Optimal variables $\hat {\bm{\phi}} ^{\rm{opt}}, T^{\rm{opt}}, R^{\rm{opt}}$
\FOR{$t=1$; ${t} \le t_{\rm{max}}$; $t++$}
\STATE Compute ${w^{{\rm{opt}}}}(t),{\mu ^{{\rm{opt}}}}(t)$ according to Eq. (\ref{Eq15}) and Eq. (\ref{Eq18});
\FOR{$k=1$; ${k} \le k_{\max}$; $k++$}
\STATE Adopt one-dimensional search to obtain $T(k)$ according to Eq. (\ref{Eq22});
\STATE For given ${w^{{\rm{opt}}}}(t),{\mu ^{{\rm{opt}}}}(t),T(k)$, solve \textbf{P(1.b)} and record the optimal $F({\bm{\phi}}(k),T(k))$;
\ENDFOR
\STATE $\{ {T^{opt}}(t),{{\bm{\phi}} ^{{\rm{opt}}}}(t)\}  = \mathop {\arg \max }\limits_{T(k),{\bm{\phi}} (k)} F(k)$;
\STATE Obtain the feasible phase ${\hat {\bm{\phi}} ^{{\rm{opt}}}}(t)$ according to Eq. (\ref{Eq32}) and compute $R^{\rm{opt}}(t)$ according to Eq. (\ref{Eq21}.b) and (c);
\STATE Update $F(t) = F({R^{{\rm{opt}}}},{T^{{\rm{opt}}}}),{\bm{\phi}}  = {{\hat {\bm{\phi}} }^{{\rm{opt}}}}(t)$;
\IF{$t$ reaches to $t_{\rm{max}}$ or {$F(t) - F(t - 1) \le ero$}}
\STATE end loop;
\ENDIF
\ENDFOR
\end{algorithmic}
\end{algorithm}

\subsection{Proposed Algorithm for \textbf{(P2)}}
As shown in Fig. \ref{Fig3}(b), \textbf{(P2.1)} can be decoupled into two independent modules, e.g., the joint optimization of $R,T$ and the optimization of $\bar C({\bm{\phi}} )$. In the following descriptions, we present the process of sub-modules optimization, respectively.

\subsubsection{Optimizing Transmit Rate $R$ and Time Allocation $T$}
For given $\bar C({\bm{\phi}} )$, \textbf{(P2.1)} can be simplified as
\begin{subequations}\label{Eq33}
\begin{align}
\textbf{(P2.a)}~~
&\mathop {\max }\limits_T \mathop {\max }\limits_R {\rm{min \{ R}}_1^{\rm{U}}{T_2},R_2^{\rm{U}}{T_1},R_1^{\rm{D}}{T_1},R_2^{\rm{D}}{T_2}\} \\
{\rm{s}}{\rm{.t}}{\rm{. }}~
&R_i^{\rm{U}}{T_i} \le R_j^{\rm{D}}{T_j},i \ne j\\
&R \le \bar C({\bm{\phi}} ),\\
&{T_1} + {T_2} = 1,
\end{align}
\end{subequations}
where the optimization of $T$ can still adopt the one-dimension search method as showed in Eq. (\ref{Eq22}). Moreover, for given $T$, the optimization of $R$ becomes a multi-objective linear programming problem, which can be easily solved via linear programming toolbox.

\subsubsection{Optimizing Channel Capacity $\bar C({\bm{\phi}} )$}
Corresponding to the expression of objectives, we rewrite the constraint Eq. (\ref{Eq33}.c) as
\begin{eqnarray}\label{Eq34}
R \le \bar C \buildrel \Delta \over = \left\{
\begin{aligned}
&{R_1^{\rm{U}}{T_2} \le \bar C_1^{\rm{U}}{T_2},}\\
&{R_2^{\rm{U}}{T_1} \le \bar C_2^{\rm{U}}{T_1},}\\
&{R_1^{\rm{D}}{T_1} \le \bar C_1^{\rm{D}}{T_1},}\\
&{R_2^{\rm{D}}{T_2} \le \bar C_2^{\rm{D}}{T_2}.}
\end{aligned}
\right.
\end{eqnarray}

Using the duality between the constraint and the objective, the sub-module of optimizing $\bar C({\bm{\phi}} )$ for given $R,T$ can be equivalent to
\begin{subequations}\label{Eq35}
\begin{align}
\textbf{(P2.b)}~~
&\mathop {\max \min }\limits_{\begin{array}{*{20}{c}}
{w,\mu }\\
{\bm{\phi}}
\end{array}} \{ \bar C_1^{\rm{U}}{T_2},\bar C_2^{\rm{U}}{T_1},\bar C_1^{\rm{D}}{T_1},\bar C_2^{\rm{D}}{T_2}\} \\
{\rm{s}}{\rm{.t}}{\rm{. }}~
&\left| {{{\bm{\phi}} _n}} \right| \le 1,\forall n = 1,...,N.
\end{align}
\end{subequations}

It can be observed that \textbf{(P2.b)} is equivalent to \textbf{(P1.2)} for given $T$ and thus, ${\bar C^{{\rm{opt}}}}$ of \textbf{(P2.b)} can be directly obtained by the result of Alg. {\ref{Alg1}}.

The details of the proposed algorithm for \textbf{(P2)} are shown in Alg. \ref{Alg2}.

\begin{algorithm}[t]
\caption{ The proposed iterative optimization algorithm for \textbf{(P2)}}
\label{Alg2}
\begin{algorithmic}
\REQUIRE ~~\\
Initial parameters ${\bm{\phi}}  = {{\bm{\phi}} _0}$
\ENSURE  ~~\\
Optimal variables $\hat {\bm{\phi}} ^{\rm{opt}}, T^{\rm{opt}}, R^{\rm{opt}}$
\FOR{$t=1$; ${t} \le t_{\rm{max}}$; $t++$}
\STATE Compute $ C({\bm{\phi}} )$;
\FOR{$k=1$; ${k} \le k_{\max}$; $k++$}
\STATE Adopt one-dimensional search to obtain $T(k)$ according to Eq. (\ref{Eq22});
\STATE For given $ C({\bm{\phi}} ),T(k)$, solve \textbf{P(2.a)} and record the optimal $F(R(k),T(k))$;
\ENDFOR
\STATE $\{ {R ^{{\rm{opt}}}(t),{T^{\rm{opt}}}(t)}\}  = \mathop {\arg \max }\limits_{R(k),T(k)} F(k)$;
\STATE Substitute $T^{\rm{opt}}(t)$ into Alg. \ref{Alg1} and output ${\hat {\bm{\phi}} ^{{\rm{opt}}}}(t)$;
\STATE Update $F(t) = F({R^{{\rm{opt}}}},{T^{{\rm{opt}}}}),{\bm{\phi}}  = {{\hat {\bm{\phi}} }^{{\rm{opt}}}}(t)$;
\IF{$t$ reaches to $t_{\rm{max}}$ or {$F(t) - F(t - 1) \le ero$}}
\STATE end loop;
\ENDIF
\ENDFOR
\end{algorithmic}
\end{algorithm}

\subsection{Complexity Analysis}
According to the process of Alg. \ref{Alg1}, we firstly need to compute $w$ in Eq. (\ref{Eq15}), where the computation of $\bm{w}_i^U$ and $w_i^D$ requires the complexity of order $O({N^2} + MN + {M^3})$ and $O({N^2} + N)$, respectively. It can be observed that there exists the gap of $O({M^3})$ imposed by the operation of matrix inversion. Meanwhile, the acquisition of $\mu$ in Eq. (\ref{Eq18}) is dominated by the computation of   $e_i^{\rm{U}}$ and $e_i^{\rm{D}}$ in Eq. (\ref{Eq16}), with the complexity of order $O({N^2} + MN)$ and $O({N^2} + N)$, respectively. Then, let us assume the complexity order of solving \textbf{P(1.b)} to be $O({\Theta _1})$, which relies on the number of constraints and optimization variables. Thereby, the one-dimensional search consumes the complexity of order $O({k_{\max }}{\Theta _1})$. Thus, the total complexity of Alg. \ref{Alg1} is limited to the order of $O({t_{\max }}({N^2} + MN + {M^3} + {k_{\max }}{\Theta _1}))$.

The process of Alg. \ref{Alg2}, compared to Alg. \ref{Alg1}, requires the additional submodule of optimizing transmit rate, due to the introduction of delay constraints. Considering the identical mathematical form of \textbf{P(1.b)} and \textbf{P(2.a)}, we denote the complexity order of solving \textbf{P(2.a)} by $O({k_{\max }}{\Theta _2})$. Based on the complexity analysis of Alg. \ref{Alg1}, the total complexity order of Alg. \ref{Alg2} can be given by $O({t_{\max }}({N^2} + MN + {M^3} + {k_{\max }}{\Theta _2} + {\Theta _1}))$. Since the cloud-management structure is adopted in our system, the main computational tasks may be carried out at the cloud, which can control the computation delay at a tolerable level.

\section{Numerical Simulation}

\subsection{Simulation Environment and Parameter Setting}
To distinguish medium 1 and 2, we assume that D1 works in the sub-6 GHz frequency band with frequency $f_1$ and D2 employs mmWave with frequency $f_2$. In addition, the difference between the transmit power of D1 and D2 is determined by
\begin{eqnarray}\label{Eq36}
\begin{aligned}
\frac{{{P_1}}}{{\sigma _1^2}} = \frac{{{P_2}}}{{\sigma _2^2}}.
\end{aligned}
\end{eqnarray}
Denoting the power spectral density of AWGN by $N_0$ identically, the average noise power is given by $\sigma_1^2 = B_1 N_0$ and $\sigma_2^2 = B_2 N_0$, respectively. Substituting the above expressions into (\ref{Eq36}), we have
\begin{eqnarray}\label{Eq36-1}
\begin{aligned}
{P_1} = \frac{{{B_1}}}{{{B_2}}}{P_2} \buildrel \Delta \over = P.
\end{aligned}
\end{eqnarray}

In the channel model, the large-scale fading adopts the log-normal pathloss model, given by
\begin{eqnarray}\label{Eq37}
\begin{aligned}
h_{ij}^{{\rm{large - scale}}} =&  - 20\log (4\pi {d_0}/\lambda_i ) - 10\eta \log ({d_{ij}}/{d_0}) + \\ &G_i~\rm{(dB)},
\end{aligned}
\end{eqnarray}
where $\lambda_i $ indicates the wavelength of medium $i$ and $\eta $ is the pathloss exponent factor. The distance between the transmitter $i$ and the receiver $j$ is denoted by $d_{ij}$, and $d_0$ is the reference distance of the free-space pathloss, $G_i$ represents the total antenna gains of medium $i$ including the device, the RIS and the AP. In addition, the ratio of the LI-plus-noise power $\tilde \sigma _i^2 $ to AWGN is 1.1.

Moreover, the small-scale fading obeying the Rician distribution, is expressed as
\begin{eqnarray}\label{Eq38}
\begin{aligned}
h_i^{{\rm{small - scale}}} =& \sqrt {{K_i}/({K_i} + 1)} \exp (1{\rm{i}}*\theta ) + \\
&\sqrt {1/({K_i} + 1)} \beta ,
\end{aligned}
\end{eqnarray}
where ${K_i}$ is the Rician factor of medium $i$, ranging from 8 to 12 dB for the mmwave channel, $\theta  \sim N(0,1)$ represents the random phase for the RF channel while $\theta  = 2\pi {d_{ij}}/\lambda_2 $ for the mmwave channel \cite{IshiN-2017}, $\beta  \sim \exp (1)$ represents the NLOS pathloss.

\begin{figure}[t]

    \centering
    \includegraphics[width=0.3\textwidth]{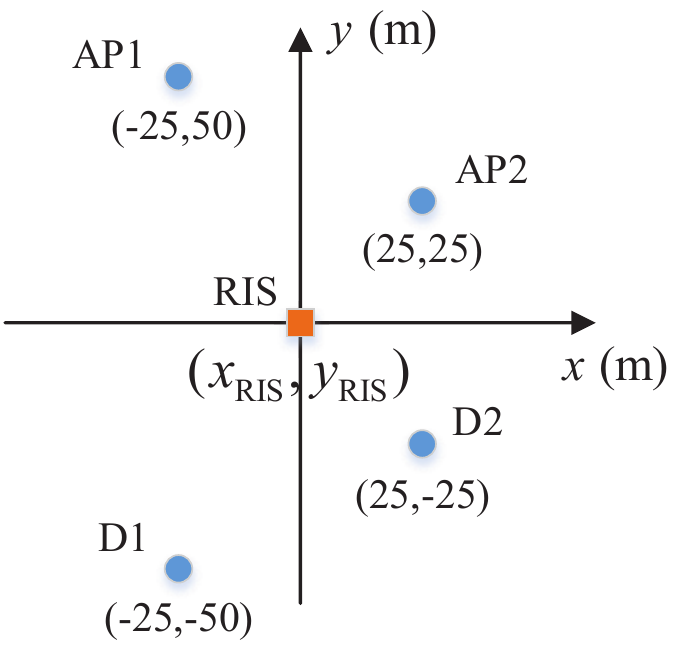}
    \caption{ The cartesian coordinate of geographical locations.}
    \label{Fig4}
\end{figure}

The geographical locations of objectives are indicated by the cartesian coordinate as shown in Fig. \ref{Fig4}, where the locations of devices and the APs are fixed while the location of the RIS is adjustable, denoted by $({x_{{\rm{RIS}}}},{y_{{\rm{RIS}}}})$. The other parameters are shown in Table \ref{Tab1} unless otherwise specified. The following simulation results are obtained by the statistical average of random implementation over 500 times.

\begin{table*}[t]
\centering
\caption{Parameter Setting}
\label{Tab1}
\begin{tabular}{|c|c|c|c|c|c|c|c|c|c|c|}  % {lccc} 表示各列元素对齐方式，left-l,right-r,center-c
\hline
Parameter & $f_1,f_2$ & $B_1, B_2 $ & $ G_1, G_2$ & $K_1, K_2$ & $M$ & $N$ & $d_0$ & $\eta $ & $\rho_{\rm{SI}}$ & $N_0$ \\ \hline %${{\rho _{\rm{SI}}}}$ \\ \hline
Value & 2.4,30 GHz & 10,100 MHz & 10,20 dB & 5,10 dB & 4 & 16 & 1 m & 2.2 & 0.5 & -174 dBm/Hz \\ \hline
\end{tabular}
\end{table*}

\begin{figure}[t]
    \centering
    \includegraphics[width=0.5\textwidth]{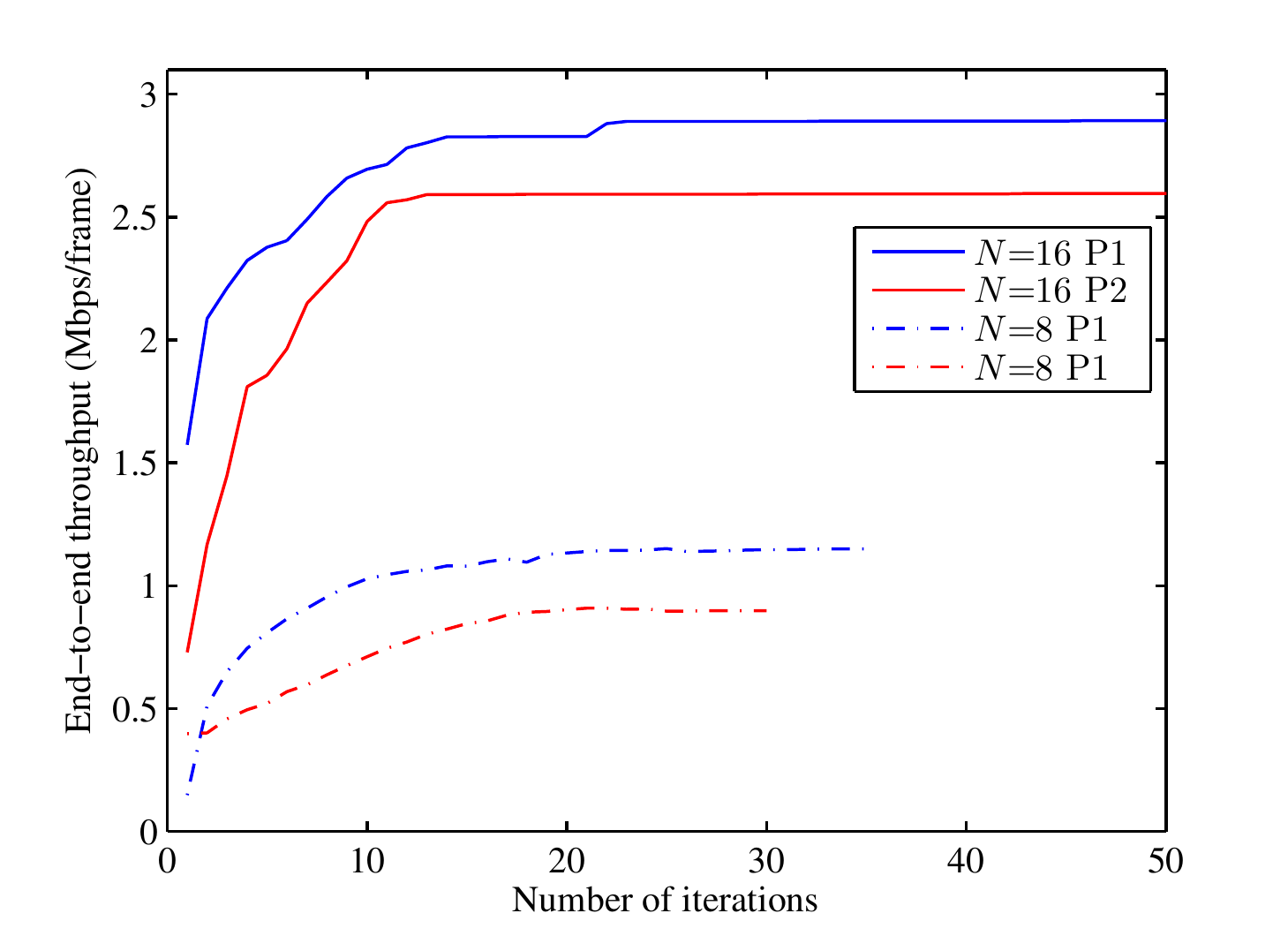}
    \caption{ The algorithm convergence of P1 and P2 versus $N$.}
    \label{Fig5}
\end{figure}

\subsection{Convergence Analysis}

Since the objective function value increases at each iteration, i.e., $F(t + 1) \ge F(t)$ for all $t$, convergence of the proposed iterative optimization algorithm can be guaranteed. Figure. \ref{Fig5} demonstrates the convergence of both the proposed algorithm 1 (P1) and 2 (P2) with $N = 8$ and $N=16$. As shown in Fig. \ref{Fig5}, the end-to-end throughput can reach a stable value through no more than 50 iterations. As seen by comparing the curves of $N = 8$ and $N=16$, less $N$ value leads to a faster convergence but a lower throughput, which verifies the effectiveness of the RIS in our scenario. Moreover, the difference between the convergence value of the proposed algorithm 1 and 2 indicates the influence of delay constraints to the end-to-end throughput.

\subsection{Benchmark: AP-Based Transmission Scheme for RIS-Aided Full-Duplex Communications}
The simulation results in \cite{HuangC-2018} and \cite{HuangC-2019} verified that the introduction of RIS can improve the EE performance compared to the cases without RIS and with an amplify-and-forward relay, respectively. In addition, since the hardware cost and circuit power of active RF chains are higher than those of the phase-shift controller \cite{TangW-2019}, it is reasonable to expect that the proposed RIS-aided reflection/transmitter scheme has an advantage of low cost over either a conventional AF relay or an MIMO transmitter. In the following, we consider an RIS-aided full-duplex two-way communication scheme \cite{PengZ-2021} as a benchmark that adopts the AP-based transmission scheme where downlink signals are transmitted from the AP as a conventional transmitter rather than the proposed RIS-based transmitter. For a fair comparison, the positions of nodes and channel realizations are set to be identical in both AP-based and RIS-based schemes. As a metric of performance comparison, the expressions of EE in the two schemes are derived as follows.

In the RIS-based scheme, the total amount of power consumption is given by
\begin{eqnarray}
\begin{aligned}
P_{{\rm{Total}}}^{{\rm{RIS - based}}} = \sum\limits_{i = 1,2} {({P_{{D_i}}} + P_{{D_i}}^{\rm{c}})}  + NP_{{\rm{RIS}}}^{\rm{c}},
\end{aligned}
\end{eqnarray}
where $P_{{D_i}}^{\rm{c}}$ represents the circuit dissipated power of node $D_i$, $P_{{\rm{RIS}}}^{\rm{c}}$ denotes the dissipated power of each RE. Thereby, the EE of the proposed RIS-based scheme can be expressed as
\begin{eqnarray}
\begin{aligned}
{\rm{E}}{{\rm{E}}_{{\rm{RIS - based}}}} = \frac{{\min \{ {S_{1 \to 2}},{S_{2 \to 1}}\} }}{{P_{{\rm{Total}}}^{{\rm{RIS - based}}}}}.
\end{aligned}
\end{eqnarray}

In the AP-based scheme, after experiencing the process of demodulation, information exchange between the APs and modulation, the APs transmit downlink signals and utilize the RIS for reflection towards devices. Specifically, the recovered signal after interference suppression can be expressed as
\begin{eqnarray}
\begin{aligned}
\hat y_i^{{\rm{D}},{\rm{AP - based}}} = & \sqrt {{P_{{\rm{A}}{{\rm{P}}_i}}}} {{\bm{g}}_{i,r}}{\rm{diag}}({\bm{\phi}} ){{\bm{g}}_{{\rm{A}}{{\rm{P}}_i},r}}x_j^l + \\
& \sqrt {{\rho _{{\rm{SI}}}}{P_{{D_i}}}} {{\bm{g}}_{i,r}}{\rm{diag}}({\bm{\phi}} ){{\bm{h}}_{i,r}}x_i^l + {\tilde n_i},
\end{aligned}
\end{eqnarray}
where ${P_{{\rm{A}}{{\rm{P}}_i}}}$ indicates the transmit power of the ${\rm{AP}}_i$, ${{\bm{g}}_{{\rm{A}}{{\rm{P}}_i},r}} \in {\mathbb{C}^{N \times 1}}$ represents the channel coefficients between the ${\rm{AP}}_i$ and the RIS.

In comparison with (\ref{Eq6}) in the RIS-based scheme, the downlink received signal in the AP-based scheme replaces $P_i$, ${{\bm{h}}_{i,r}}$ by the transmit power of the AP ${P_{{\rm{A}}{{\rm{P}}_i}}}$ and the channel coefficient ${{\bm{g}}_{{\rm{A}}{{\rm{P}}_i},r}}$ between the APs and the RIS, respectively. Thereby, the downlink channel capacity in the RIS-based scheme can be derived by
\begin{eqnarray}\label{Eq43}
\begin{aligned}
C_i^{\rm{D}}=& {B_i}\log (1 + \frac{{{P_{{\rm{A}}{{\rm{P}}_i}}}{{\left| {{{\bm{g}}_{{\rm{A}}{{\rm{P}}_i},r}} {\bm{\phi}} {{\pmb{h}_{i,r}}}} \right|}^2}}}{{{\rho _{\rm{SI}}}{P_{{\rm{A}}{{\rm{P}}_i}}}{{\left| {{{\bm{g}}_{{\rm{A}}{{\rm{P}}_i},r}} {\bm{\phi}} {{\pmb{h}_{i,r}}}} \right|}^2} + \tilde \sigma _i^2}}).
\end{aligned}
\end{eqnarray}

Besides the mentioned difference in channel capacity, the AP-based scheme dispenses with the optimization of time allocation since the two media-independent APs can simultaneously transmit data without mutual interferences during the whole frame, i.e., ${T_1} = {T_2} = {T_{\max }}$. Thus, the end-to-end throughput in the AP-based scheme is given by
\begin{subequations}\label{Eq44}
\begin{align}
S_{1 \to 2}^{{\rm{AP - based}}} =& \min \{ R_1^U,R_2^D\} {T_{\max }},\\
S_{2 \to 1}^{{\rm{AP - based}}} =& \min \{ R_2^U,R_1^D\} {T_{\max }}.
\end{align}
\end{subequations}

By letting ${T_1} = {T_2} = {T_{\max }}$ and replacing the objective function of \textbf{(P2)} by (\ref{Eq44}) and the channel capacity by (\ref{Eq43}), we obtain the optimization model of the AP-based scheme, where the optimal transmit rate and RIS phase can be found via the similar iterative optimization process of the proposed algorithm for \textbf{(P2)}.

In the AP-based scheme, the total power consumption is given by
\begin{eqnarray}
\begin{aligned}
P_{{\rm{Total}}}^{{\rm{AP - based}}} =& \sum\limits_{i = 1,2} {({P_{{D_i}}} + P_{{D_i}}^{\rm{c}})}  + NP_{{\rm{RIS}}}^{\rm{c}} + \\
& \sum\limits_{i = 1,2} {({P_{{\rm{A}}{{\rm{P}}_i}}} + P_{{\rm{A}}{{\rm{P}}_i}}^{\rm{c}})}.
\end{aligned}
\end{eqnarray}
where $P_{{\rm{A}}{{\rm{P}}_i}}^{\rm{c}}$ denotes the circuit dissipated power of the ${\rm{AP}}_i$. Thereby, the EE of the AP-based scheme can be expressed as
\begin{eqnarray}
\begin{aligned}
{\rm{E}}{{\rm{E}}_{{\rm{AP - based}}}} = \frac{{\min \{ S_{1 \to 2}^{_{{\rm{AP - based}}}},S_{2 \to 1}^{_{{\rm{AP - based}}}}\} }}{{P_{{\rm{Total}}}^{{\rm{AP - based}}}}}.
\end{aligned}
\end{eqnarray}

\begin{figure}[t]
\centering
\subfigure[End-to-end throughput of the AP-based and the RIS-based scheme versus the location of the RIS.]{
\centering
\includegraphics[width=0.48\textwidth]{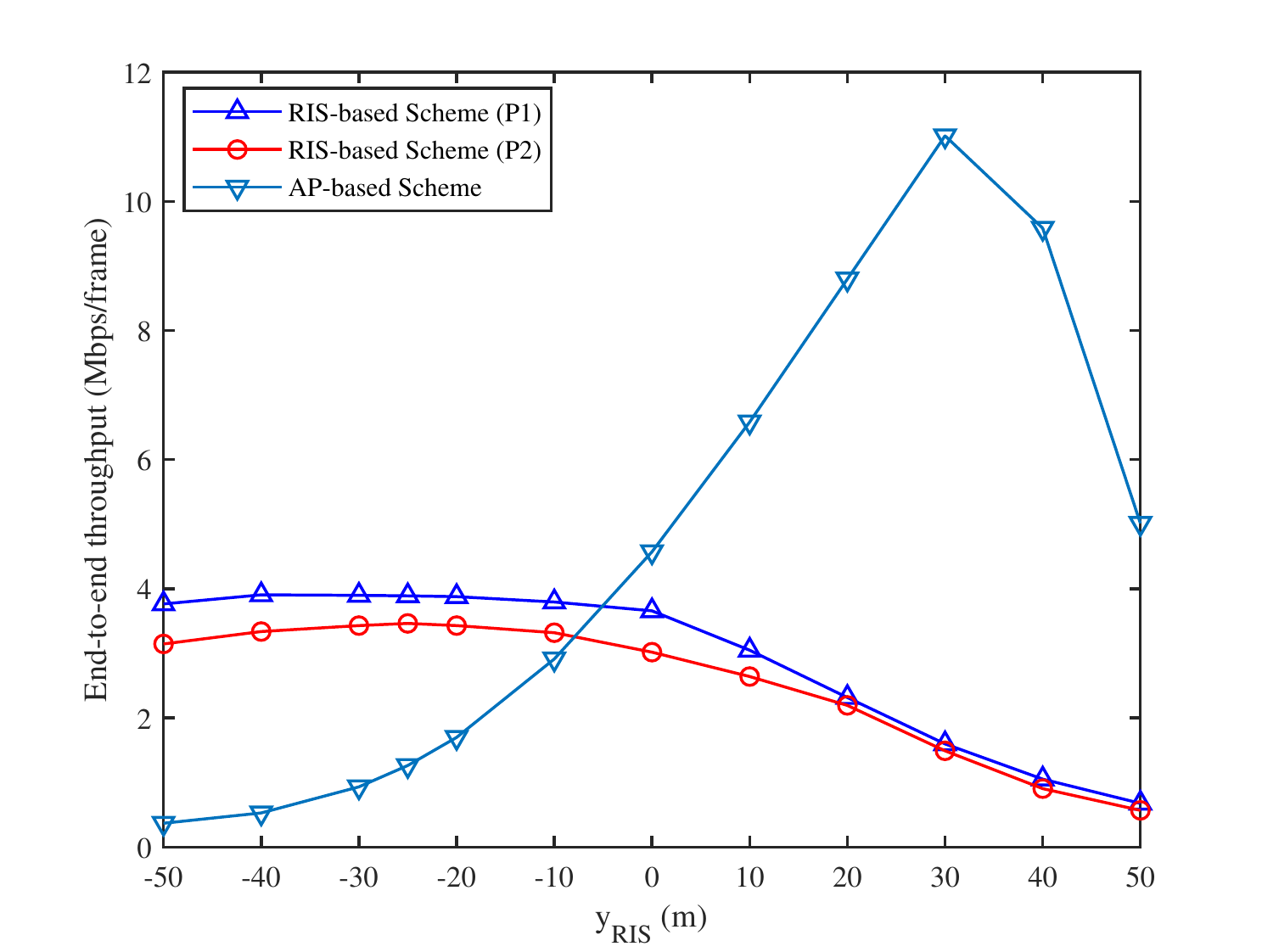}}
\subfigure[End-to-end throughput of the AP-based and the RIS-based scheme versus the location of the RIS.]{
\centering
\includegraphics[width=0.48\textwidth]{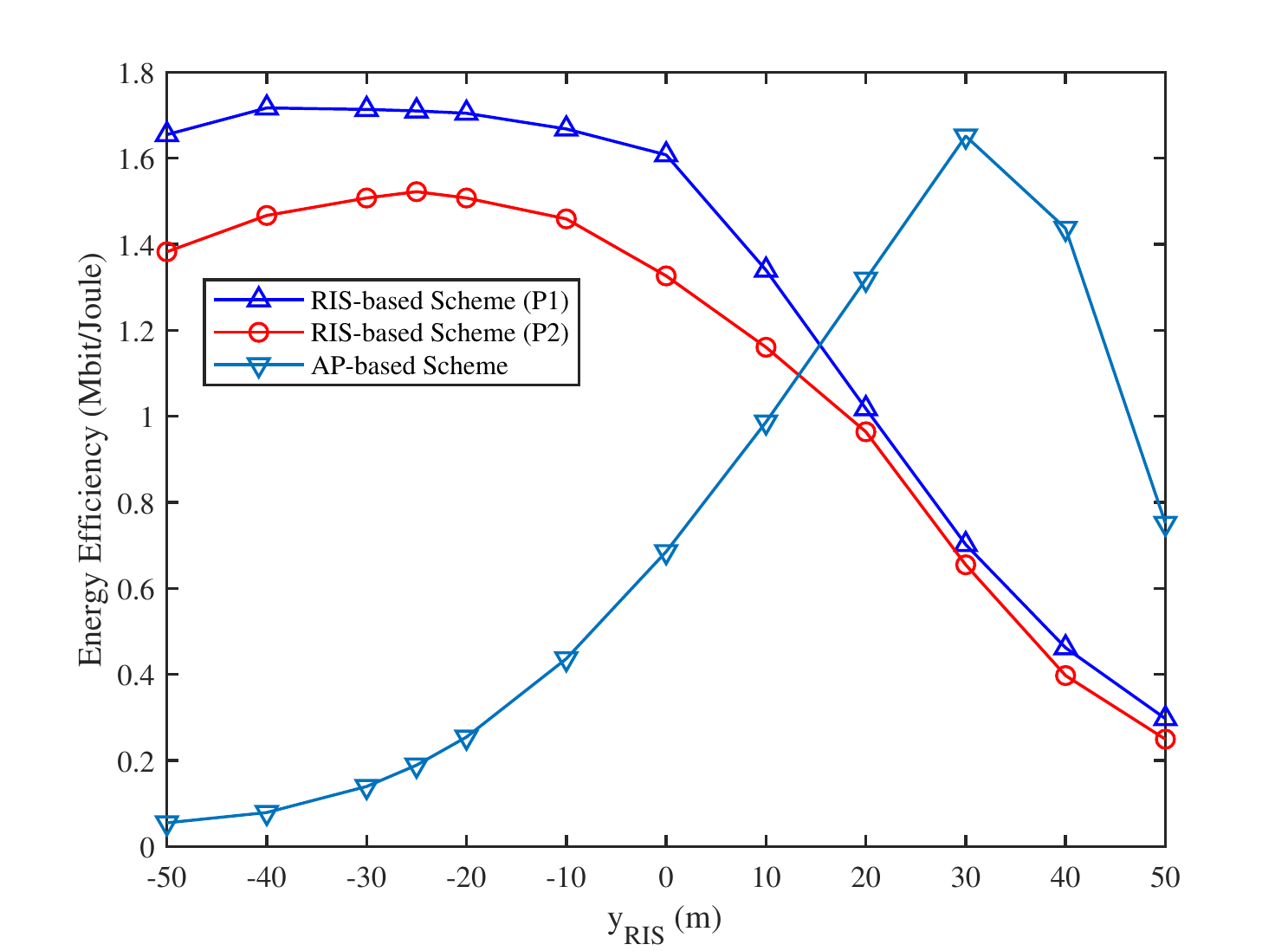}}
\caption{Performance comparison between the AP-based and the RIS-based scheme versus the location of the RIS.}
\label{Fig6}
\end{figure}

\begin{figure}[t]
\centering

\subfigure[The transmit power versus the end-to-end throughput with different $\rho_{\rm{e}}$.]{
\centering
\includegraphics[width=0.48\textwidth]{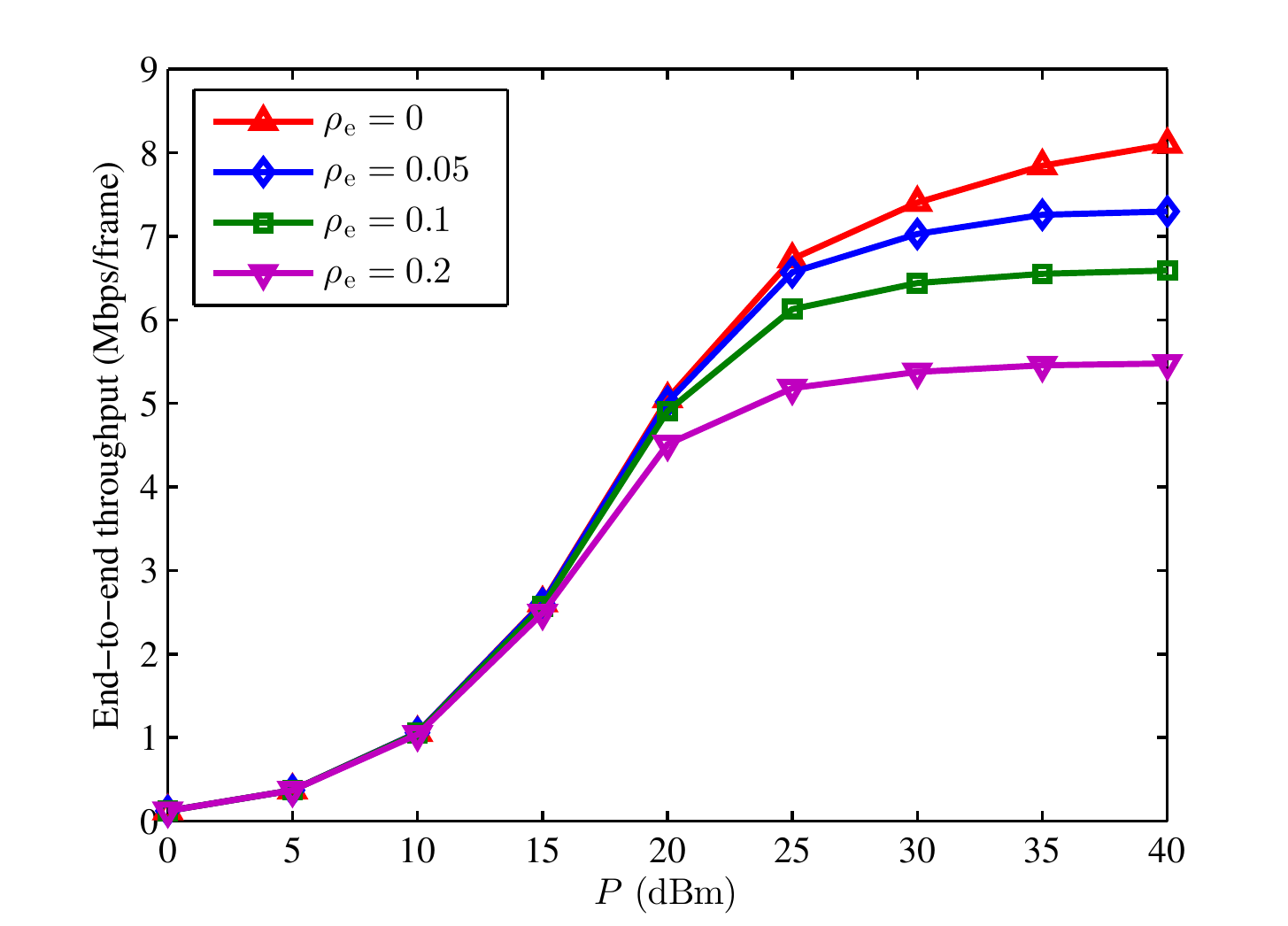} }
\subfigure[The transmit power versus the end-to-end throughput with different $\rho_{\rm{SI}}$.]{
\centering
\includegraphics[width=0.48\textwidth]{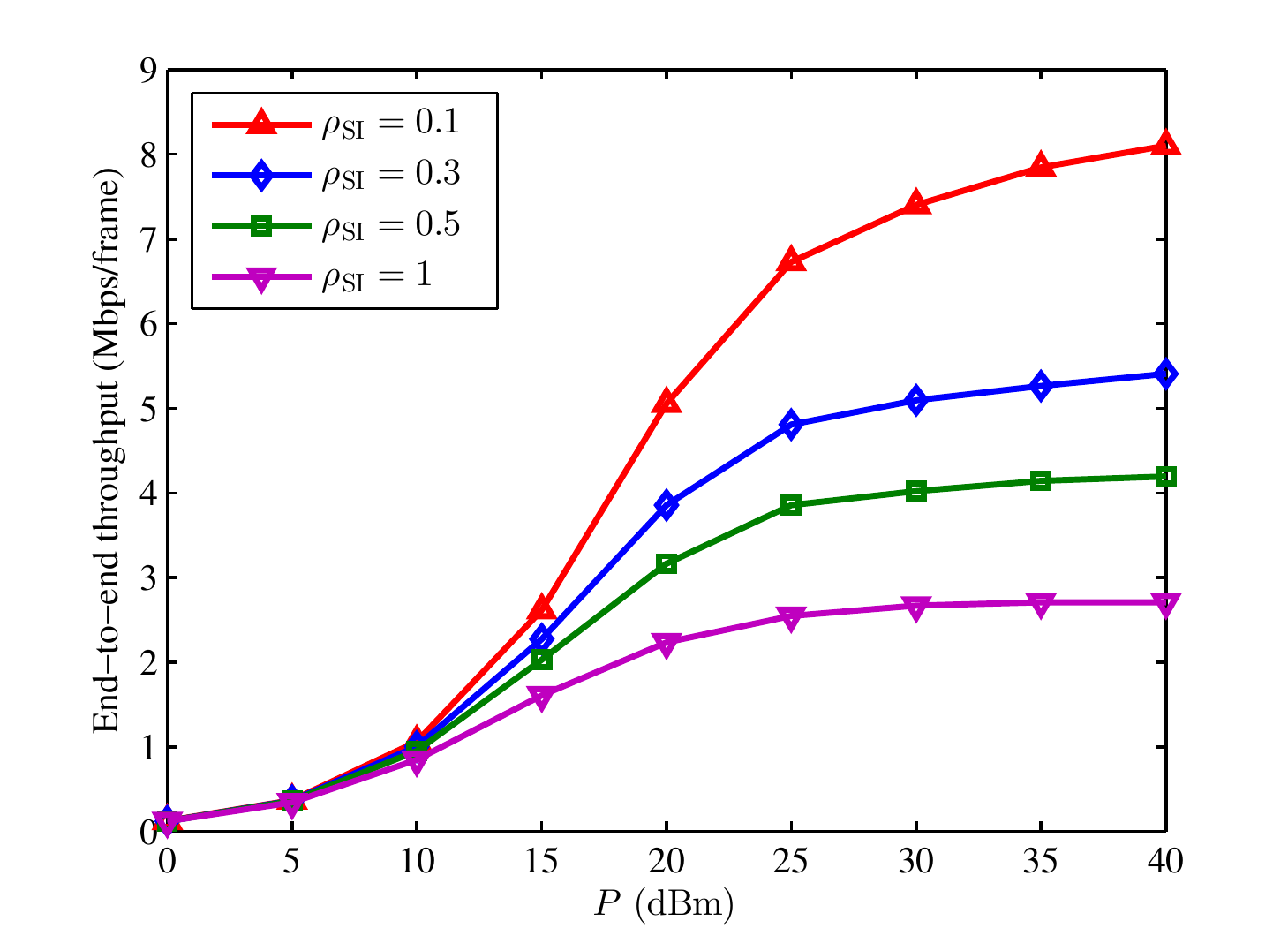} }
\caption{ The effects of ${\rho _{\rm{e}}}$ and ${\rho _{\rm{SI}}}$ on system performance.}
\label{Fig7}
\end{figure}
\subsection{Performance Comparison between the RIS-based and the AP-based Scheme}
To demonstrate the advantages of the proposed RIS-based scheme over the AP-based one, performance comparison of the end-to-end throughput and the EE in two schemes are carried out via numerical simulations. By setting $N = 16$, ${P_1} = \frac{{{B_1}}}{{{B_2}}}{P_2} = {P_{{\rm{A}}{{\rm{P}}_1}}} = \frac{{{B_1}}}{{{B_2}}}{P_{{\rm{A}}{{\rm{P}}_2}}} = 23{\rm{dBm}}$, $P_{\rm{1}}^{\rm{c}} = \frac{{{B_1}}}{{{B_2}}}P_{\rm{2}}^{\rm{c}} = 10{\rm{mW}}$, $P_{{\rm{A}}{{\rm{P}}_{\rm{1}}}}^{\rm{c}} = \frac{{{B_1}}}{{{B_2}}}P_{{\rm{A}}{{\rm{P}}_{\rm{2}}}}^{\rm{c}} = 200{\rm{mW}}$, $P_{{\rm{RIS}}}^{\rm{c}} = 5{\rm{mW}}$ \cite{ZhouG-2020}, the simulation results are shown as follows.

Figures. \ref{Fig6}(a) and \ref{Fig6}(b) compare the performance of the end-to-end throughput and the EE between the RIS-based and the AP-based downlink transmission versus the location of the RIS. As observed in Fig. \ref{Fig6}(a), when the RIS is geographically near to the devices, the throughput and EE performance of the RIS-based downlink transmission are superior to that of the AP-based downlink transmission. As the RIS is far away from the devices but closer to the AP, the throughput and EE performance of the AP-based downlink transmission exceed that of the RIS-based scheme. Although suffering from the part loss of the end-to-end throughput when the RIS is far away from the devices, the saved energy in the RIS-based downlink transmission can be used for other tasks at the APs, such as homogeneous media communications, potentially leading to a higher EE than that as shown in Fig. \ref{Fig6}(b).
Based on the above observations, the RIS-based transmission scheme is used for low-cost and high EE in our model. In the case of dynamic users, a hybrid RIS-based and AP-based scheme may be more beneficial to maximize the effects of the proposed design, which is worth further researches in future works.

\begin{figure}[t]
\centering

\subfigure[The channel capacity versus the transmit power.]{
\centering
\includegraphics[width=0.45\textwidth]{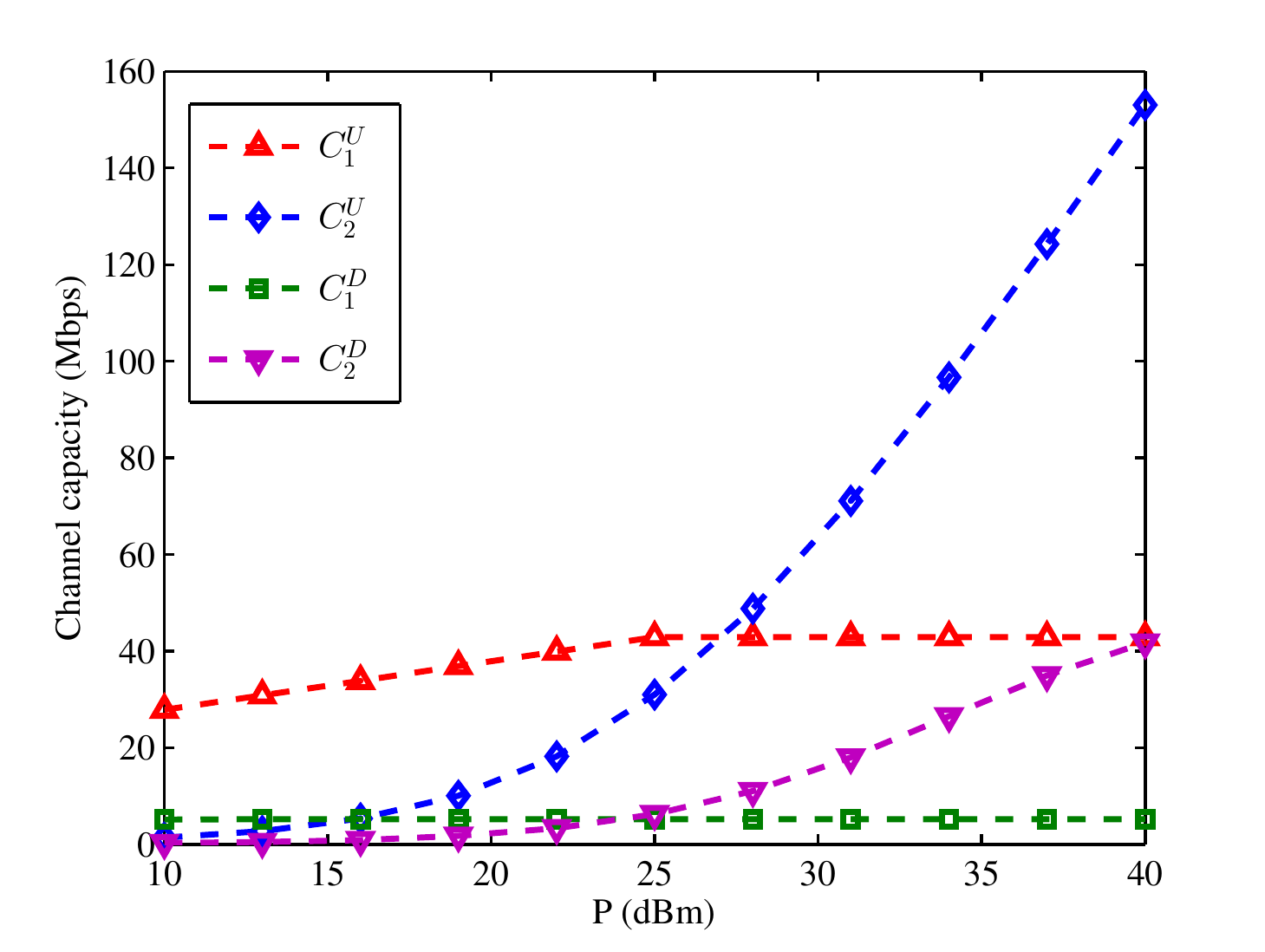} }
\subfigure[Time allocation versus the transmit power.]{
\centering
\includegraphics[width=0.48\textwidth]{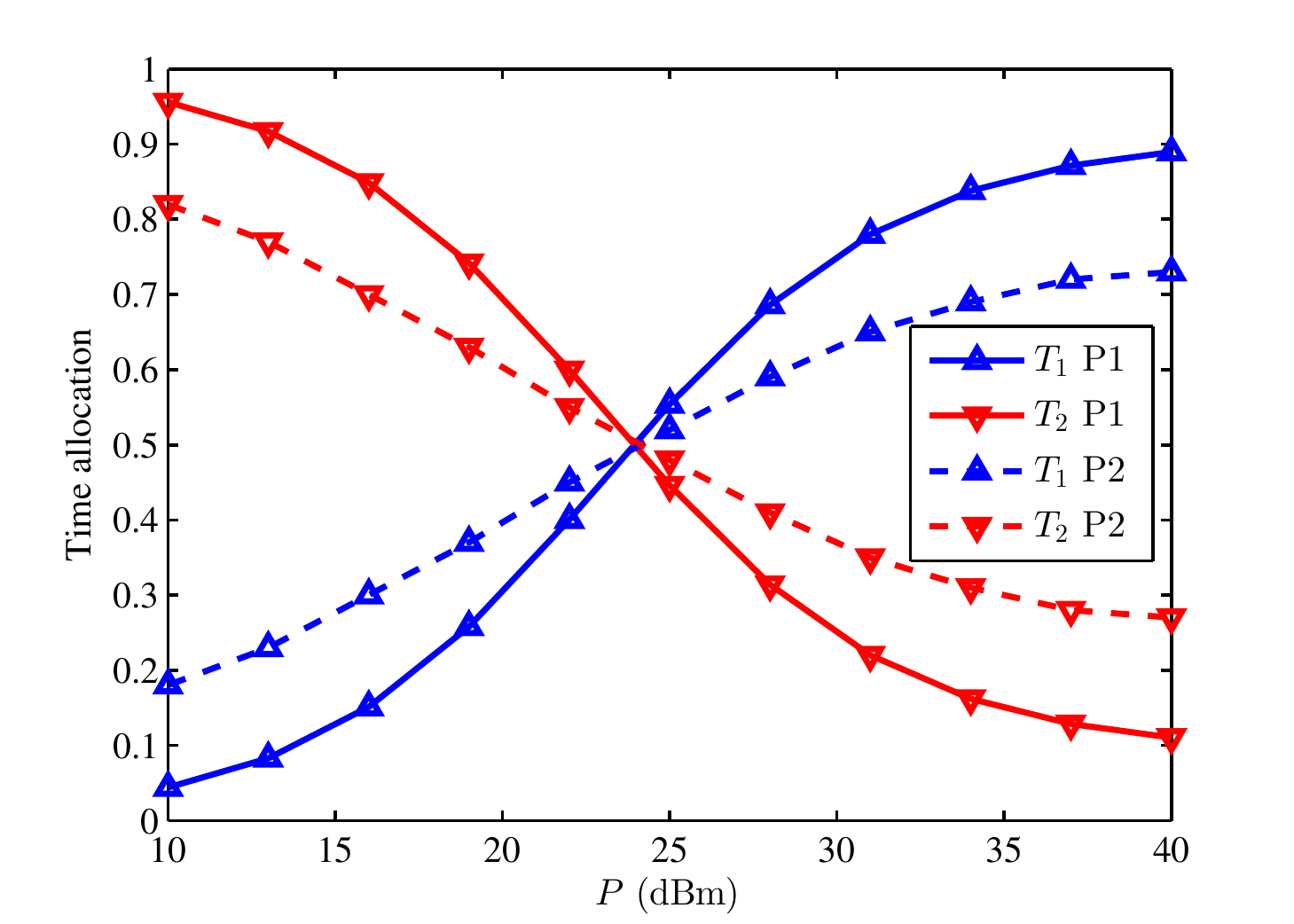} }
%\caption{  (T,WPR) mode against J attack}
%\label{Fig:6.b}
\subfigure[The transmit rate of P1 versus the transmit power.]{
\centering
\includegraphics[width=0.48\textwidth]{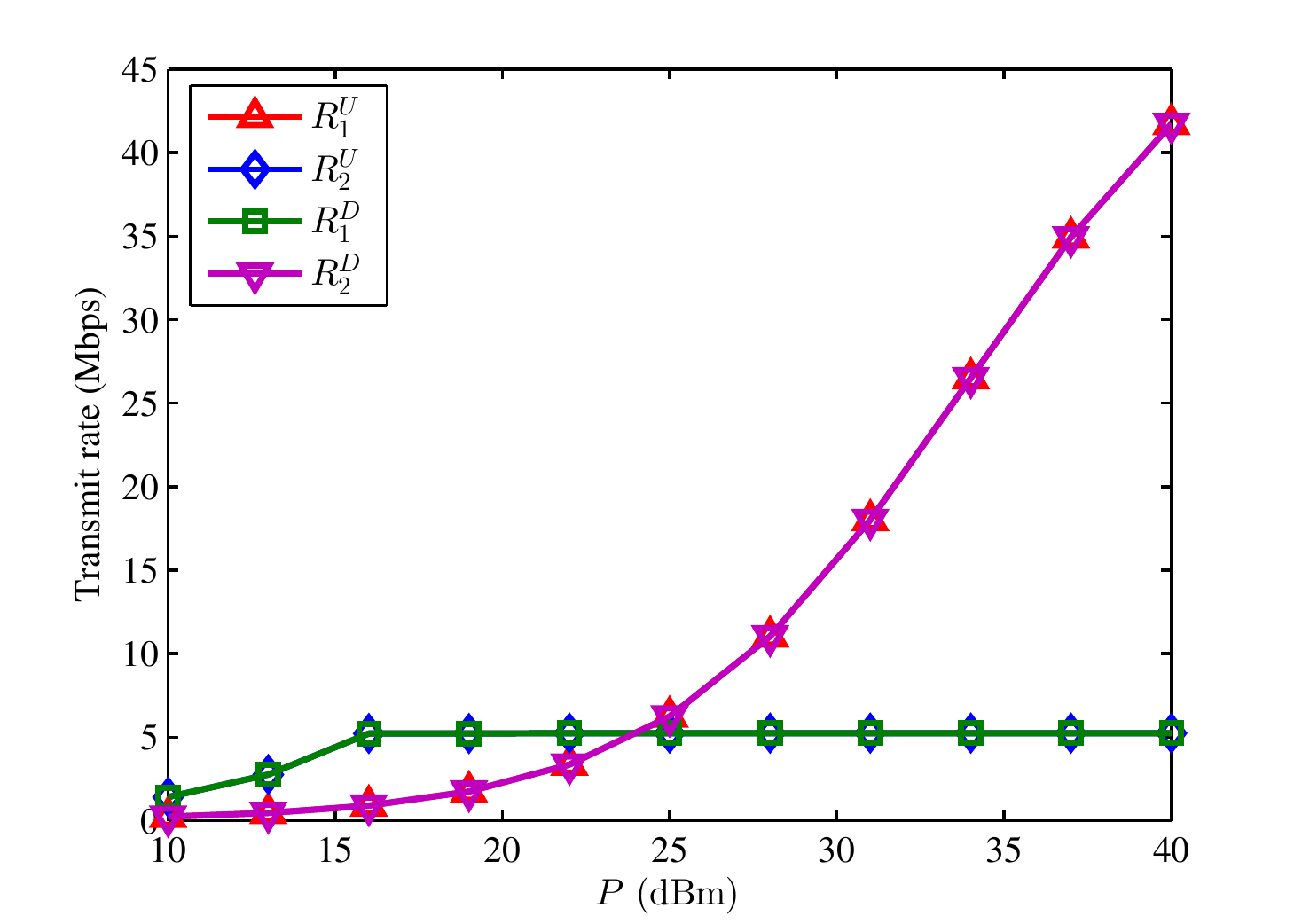} }
%\end{figure*}
%\begin{figure}
\centering
\subfigure[The transmit rate of P2 versus the transmit power.]{
\centering
\includegraphics[width=0.45\textwidth]{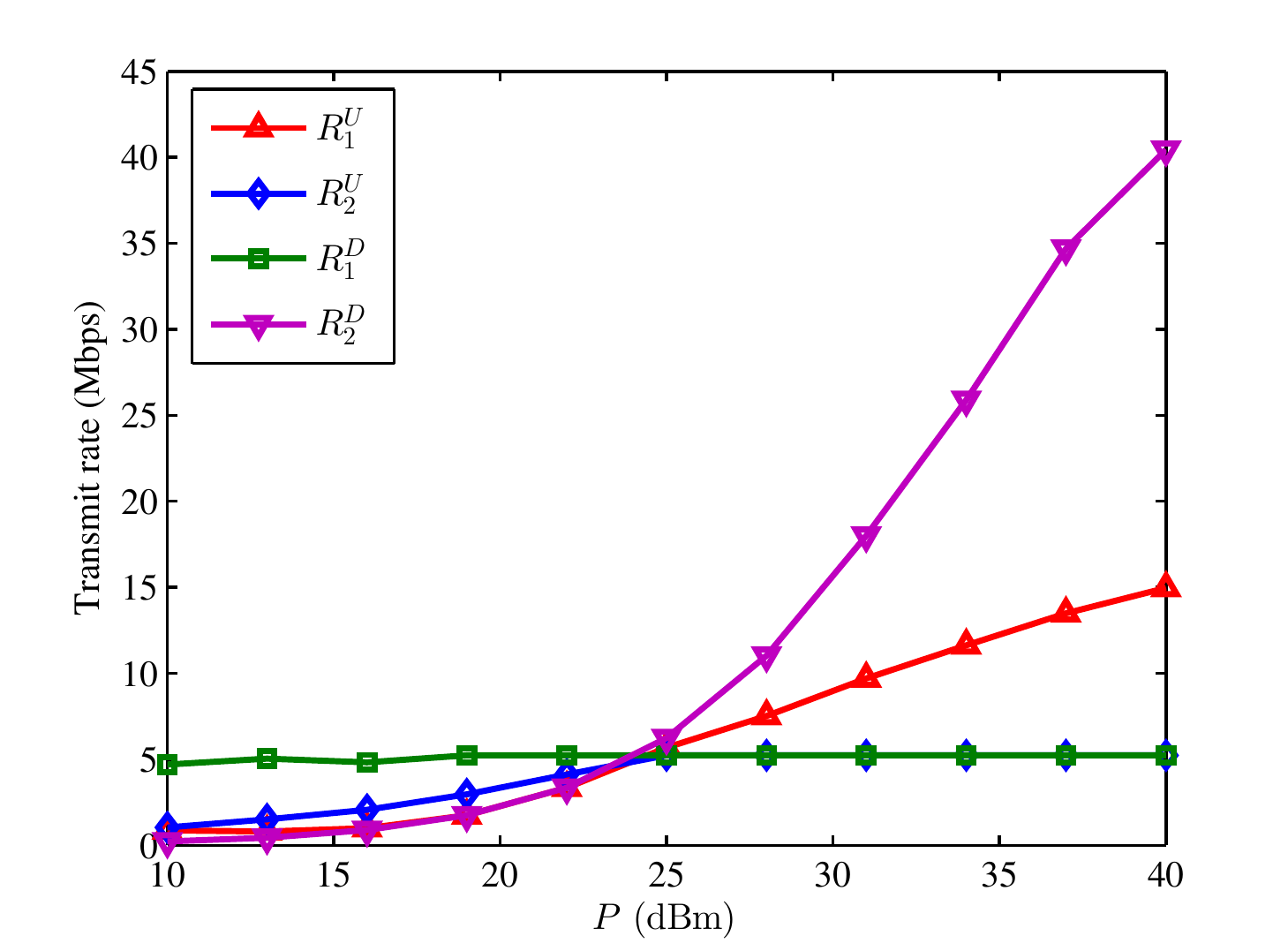} }
\caption{The influence of the transmit power on the tuning parameters.}
\label{Fig8}
\end{figure}
\subsection{The Influence of channel estimation error and SI on System Performance}
In the case of the imperfect CSI, channel estimation based on the maximum likelihood (ML) is modeled as \cite{ZhouG-2020-2}
\begin{eqnarray}\label{Eq39}
\begin{aligned}
\bm{H} = \hat {\bm{H}} + {{\rho _e}} \tilde {\bm{H}},
\end{aligned}
\end{eqnarray}
where $\bm{H}$ is the physical channel, $\hat {\bm{H}}$ is the estimated channel, ${\rho _e}$ is the estimation error coefficient, $\tilde {\bm{H}}$ introduces the random estimation errors following the complex Gaussian distribution with zero-mean and pathloss variance denoted by $\alpha$.

By replacing the channel coefficients with the estimated values, the proposed algorithm can be extended to the case of the imperfect CSI while the increase of average noise power reflects the effects of channel estimation errors on system performance, i.e., increase from $\sigma _0^2$ to $(1 + \rho _{\rm{e}}^2{P_i}{\alpha ^{\rm{U}}_i}/\sigma _0^2)\sigma _0^2$ in the uplink receive SINRs and increase from $\sigma _i^2$ to $(1 + \rho _{\rm{e}}^2{P_i}\alpha _i^{\rm{D}}/\sigma _i^2 + \rho _{\rm{e}}^2{\rho _{{\rm{SI}}}}{P_i}\alpha _i^{\rm{D}}/\sigma _i^2)\sigma _i^2$ in the downlink receive SINRs.

To show the effects of channel estimation errors and SI on system performance, we carry out numerical simulation of the transmit power versus the end-to-end throughput with different ${\rho _{\rm{e}}}$ in Fig. \ref{Fig7}(a) and ${\rho _{\rm{SI}}}$ in Fig. \ref{Fig7}(b), respectively. As observed in Fig. \ref{Fig7}, the end-to-end throughput decreases with increasing ${\rho _{\rm{e}}}$ and ${\rho _{\rm{SI}}}$. In the case of low and medium SINRs, the effects of channel estimation errors on system performance are not comparable to that of SI, which exhibits the robustness of the proposed scheme against channel estimation errors.

\subsection{The Influence of the Transmit Power on Tuning Parameters}

\begin{figure}[t]
\centering
\includegraphics[width=0.5\textwidth]{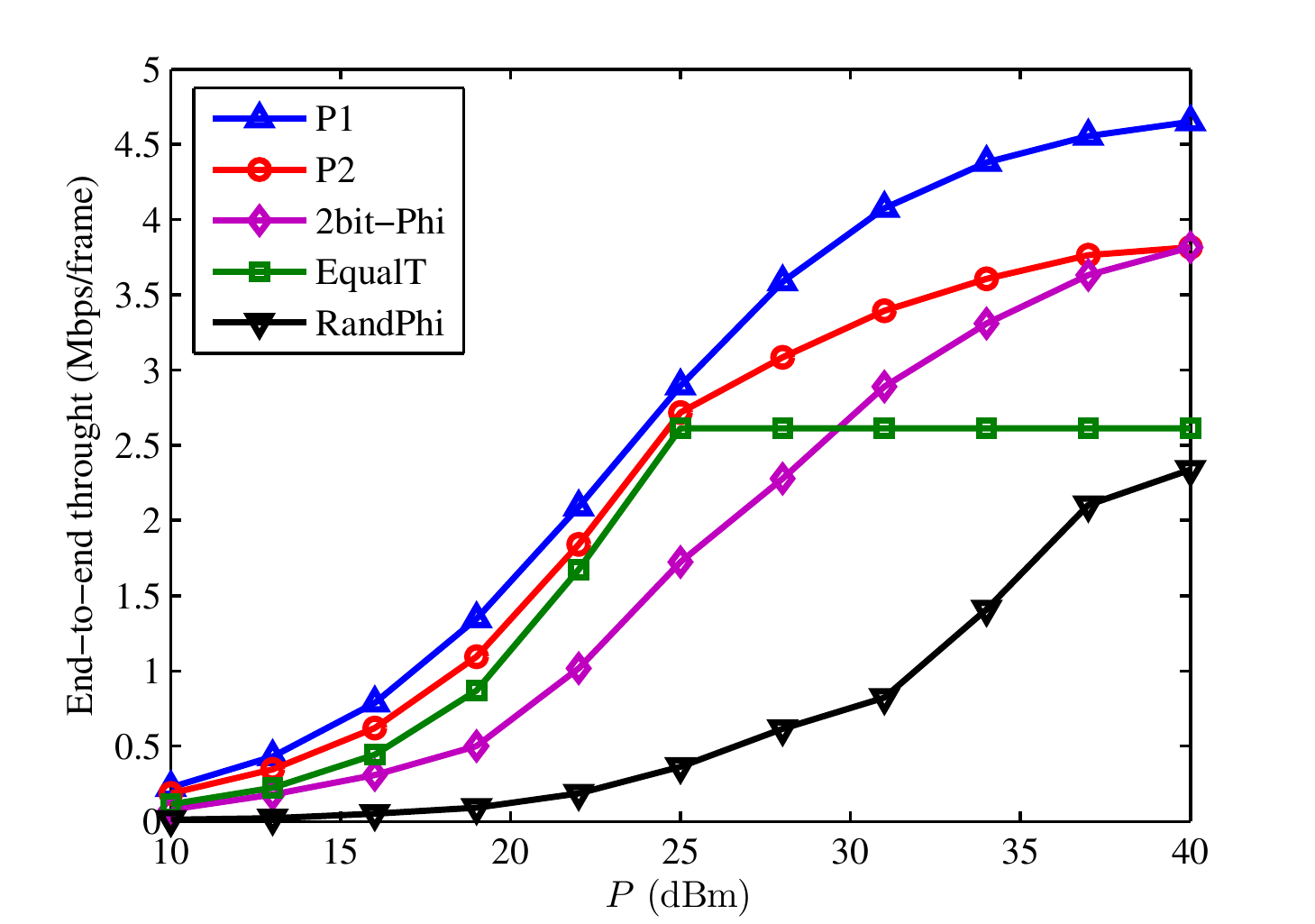}
\caption{ The end-to-end throughput of different schemes versus the transmit power $P$.}
\label{Fig9}
\end{figure}

\begin{figure}[t]
\centering

\subfigure[The channel capacity versus the RIS location $x_{\rm{RIS}}$.]{
\centering
\includegraphics[width=0.45\textwidth]{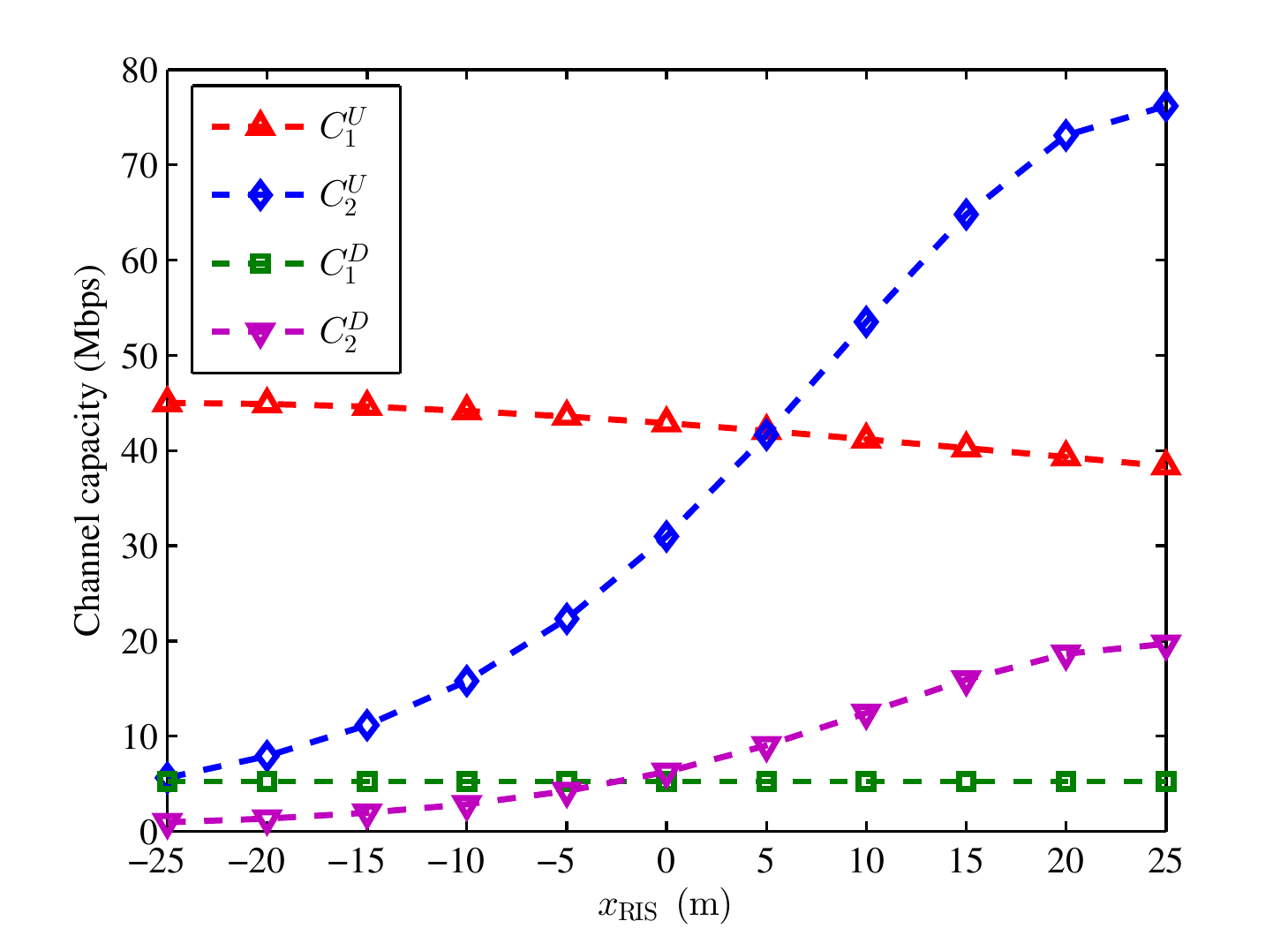} }
\subfigure[The end-to-end throughput versus the RIS location $x_{\rm{RIS}}$.]{
\centering
\includegraphics[width=0.45\textwidth]{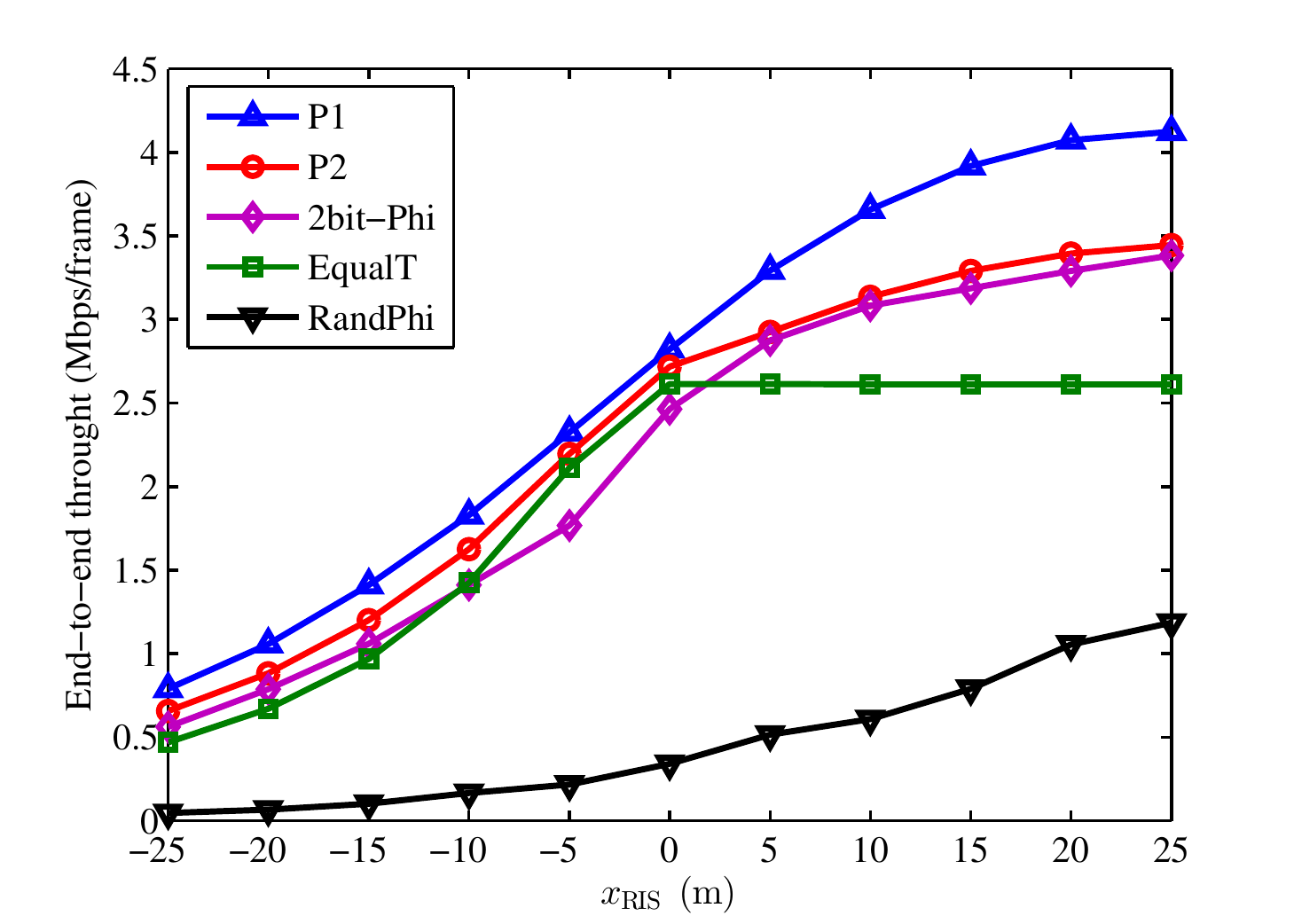} }
%\caption{  (T,WPR) mode against J attack}
%\label{Fig:6.b}
\subfigure[The channel capacity versus the RIS location $y_{\rm{RIS}}$.]{
\centering
\includegraphics[width=0.45\textwidth]{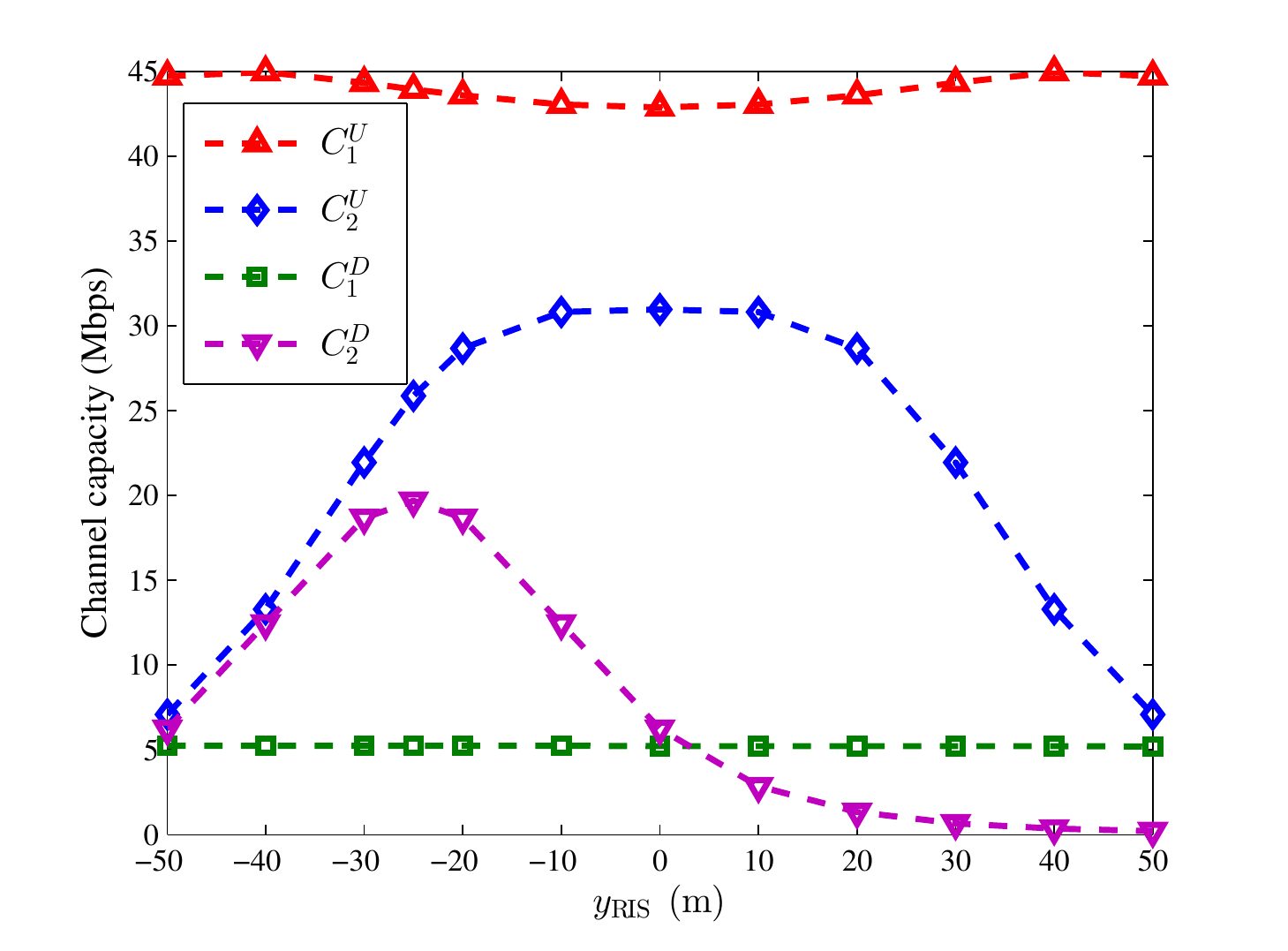} }
%\end{figure*}
%\begin{figure}
\centering
\subfigure[The end-to-end throughput versus the RIS location $y_{\rm{RIS}}$.]{
\centering
\includegraphics[width=0.45\textwidth]{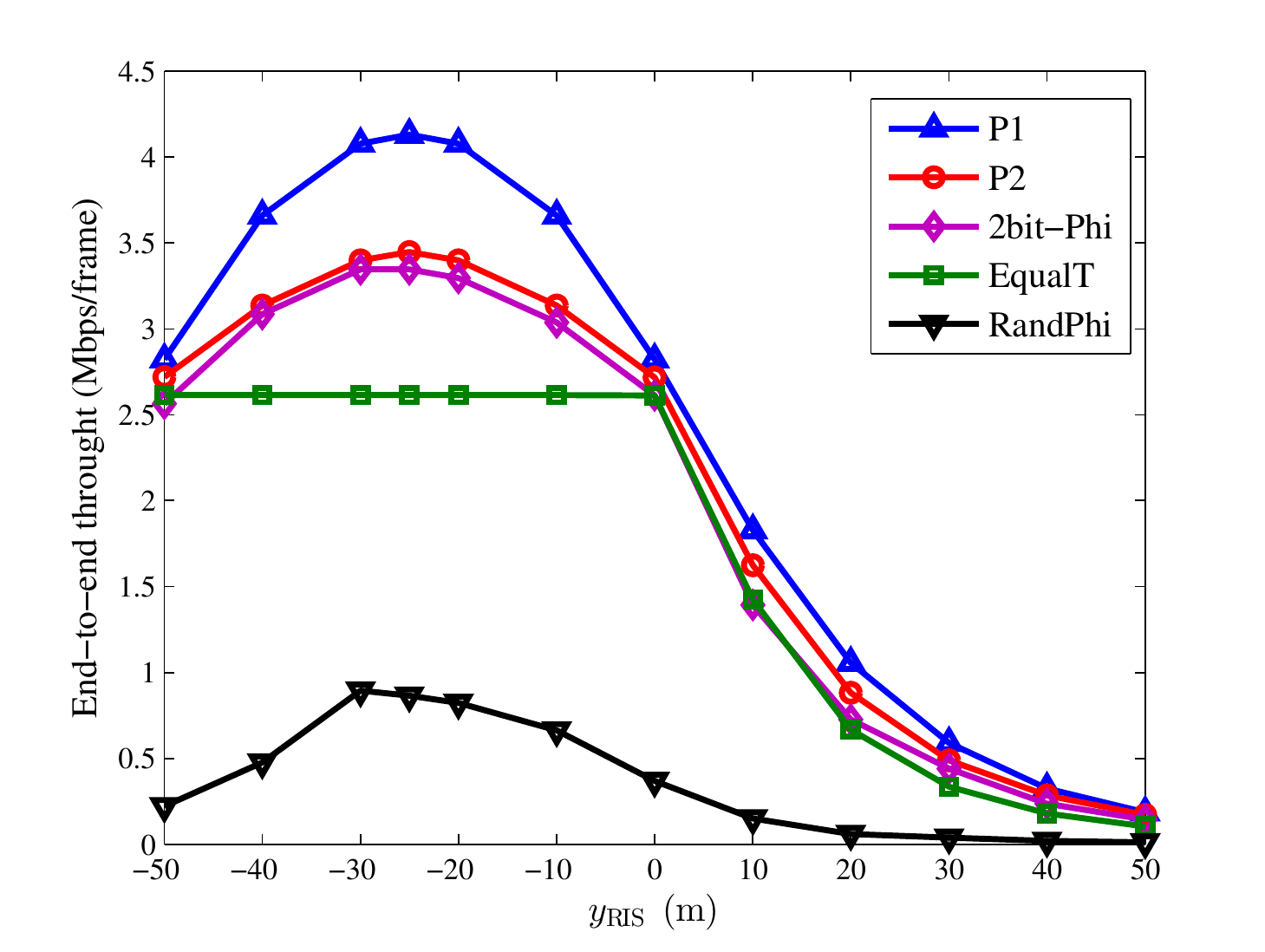} }
\caption{The influence of the RIS location on system performance of different RIS-based schemes.}
\label{Fig10}
\end{figure}

Figure. \ref{Fig8} shows the influence of the transmit power on tuning parameters of P1 and P2, including channel capacity [Fig. \ref{Fig8}(a)], time allocation [Fig. \ref{Fig8}(b)], transmit rate of P1 [Fig. \ref{Fig8}(c)] and P2 [Fig. \ref{Fig8}(d)]. As seen in Fig. \ref{Fig8}(a), the channel capacity of uplink is higher than that of downlink, owing to the lower SI and more receive antennas at the BS side compared to those at the device side. In addition, we observe that D1 achieves a higher channel capacity than D2 when $P$ is low. However, the increase of channel capacity for D2 with increasing $P$ is more significant than that for D1. This is because the channel fading of media 2 is more severe than that of media 1, but the higher $P$  can make the effect of bandwidth on capacity more significant than that of channel fading. Moreover, as shown in Fig. \ref{Fig8}(b), the difference between $T_1$ and $T_2$ for P2 is smaller than that for P1, which indicates the limitation of delay constraints to the ratio of time allocation.

According to Eq. (\ref{Eq21}.b) and (c), the transmit rates $R_i^{\rm{U}}$, $R_j^{\rm{D}}$ for P1 are identical to those determined by the minimum between uplink capacity of one device and downlink capacity of the other one. Thus, as shown in Fig. \ref{Fig8}(c), two curves for $R_i^{\rm{U}}$ and $R_j^{\rm{D}}$ overlap with each other and are both equal to the lower part between the channel capacity $C_i^{\rm{U}}$ and $C_j^{\rm{D}}$ in Fig. \ref{Fig8}(a). In addition, combined with the results of time allocation for P1 in Fig. \ref{Fig8}(b), we observe that a higher end-to-end transmit rate matches a lower preallocated slot, in order to guarantee the fairness of information exchange. By contrast, Fig. \ref{Fig8}(d) presents that the introduction of delay constraints in P2 can make a difference between $R_i^{\rm{U}}$ and $R_j^{\rm{D}}$. From the results, we notice that as the transmit power increases, the ratio of $R_2^{\rm{D}}$ to $R_1^{\rm{U}}$ increases, while the ratio of $R_1^{\rm{D}}$ to $R_2^{\rm{U}}$ decreases and finally maintains 1. According to Eq. (\ref{Eq33}.b), satisfying the delay constraints requires that the ratio of $R_j^{\rm{D}}$ to $R_i^{\rm{U}}$ is greater than the ratio of $T_i$ to $T_j$. Thus, when the ratio of $T_1$ to $T_2$ for P2 increases with increasing transmit power as shown in Fig. \ref{Fig8}(b), the transmit rate makes a corresponding adaption to meet the delay constraints.

\subsection{Baseline Schemes Based on the RIS}
To demonstrate the effectiveness of the proposed scheme, several RIS-based baseline schemes with delay constraints are introduced, respectively denoted by ``2bit-Phi'', ``EqualT'' and ``RandPhi''. Therein, the 2bit-Phi scheme discretizes the RIS phase using 2-bit uniform quantization, i.e., ${\bm{\phi}} _i^{2 - {\rm{bit}}} \in [0,\pi /2,\pi ,3\pi /2]$, and the feasible phase will be determined by minimizing $\left| {{\bm{\phi}} _i^{2 - {\rm{bit}}} - {\bm{\phi}} _i^{{\rm{opt}}}} \right|$ . The EqualT scheme will optimize the RIS phase as the proposed scheme, whereas simply equalizing the preallocated slot, i.e., ${T_1} = {T_2} = 0.5$. The RandPhi scheme adopts the random phase adjustment and the equal time allocation.

Figure. \ref{Fig9} shows the end-to-end throughput of information exchange versus the transmit power for different schemes. From the results, we observe that the proposed scheme can obtain a significant performance gain, by contrast to other benchmarks, where P1 without delay constraints achieves the best performance while the random phase scheme behaves worst. The significant difference between the EqualT and RandPhi schemes indicates the effectiveness of phase adjustments. As $P$ increases, the influence of phase adjustments on the throughput becomes quite limited, leading to a nonincreasing trend for the EqualT. However, as we noticed in the comparison of the P2 and EqualT curves, optimizing time allocation still exhibits the effectiveness for any given $P$. In addition, 2bit-Phi scheme will suffer from an at most 30\% performance loss in the throughput, compared to that of the continuous phase.

In Fig. \ref{Fig10}, we evaluate the effect of the RIS location on the channel capacity of the proposed scheme and compare the end-to-end throughput of different schemes against ($x_{\rm{RIS}}$,0) [Fig. \ref{Fig10}(a)(b)] and (0,$y_{\rm{RIS}}$) [Fig. \ref{Fig10}(c)(d)]. As seen in Fig. \ref{Fig10}(a), the channel capacity of D2 increases more than the decrease in D1 when moving the RIS from the horizontal location of D1 to that of D2. This is because medium 1 is more insensitive to the variation
of the distance than medium 2, thus the effects of distance on the system performance are mainly dominated by the distance between the RIS and D2. In Fig. \ref{Fig10}(c), when moving the RIS from the vertical location of D1 to that of D2, the channel capacity of D1 changes slightly but that of D2 exhibits a significant trend of firstly increasing and then decreasing. The maximum values for $C_2^{\rm{U}}, C_2^{\rm{D}}$ are achieved at the point of -25m and 0m, respectively. As expected, the curves of the end-to-end throughputs in Fig. \ref{Fig10}(b) and (d) have a similar trend with the channel capacity of D2 in Fig. \ref{Fig10}(a) and (c), excepted for the EqualT. The curve of the end-to-end throughput for the EqualT can be explained by comparing the minimum value between $C_1^{\rm{D}}$ and $C_2^{\rm{D}}$, since the EqualT scheme lacking of time allocation optimization only maximizes the minimum rate between them for improving the end-to-end throughput. In general, the results of performance comparison versus the RIS location in Fig. \ref{Fig10} represent a qualitative relationship similar to that depicted in Fig. \ref{Fig9}, which demonstrates the general advantage of the proposed scheme in improving the transmission performance compared to other benchmarks.

\section{Conclusion}
In summary, we proposed an RIS-aided hybrid reflection/transmitter structure, towards instantaneous information exchange in cross-media communications. In the context of the cloud-management protocol, the cross-media information exchange could be achieved at a low cost without substantially modifying the current network structure. Specifically, the RIS can perform intelligent reflection to enhance the received SNR in the full-duplex transmission meanwhile the RIS is utilized as a transmitter to reduce energy consumption in the downlink transmission. Since different media generate no mutual interferences, the uplink signals from different devices can be reflected from the RIS to their respective APs at the same time. On the other hand, the downlink signals that are modulated onto the phases of incident carrier signals can be reflected from the RIS to the destination nodes based on TDMA protocol. Through the joint optimization of the RIS phase, transmit rate and time allocation, the end-to-end throughput of information exchange achieves maximization while the feedback delay could be controlled within a tolerant level. Our simulation results revealed that the proposed scheme improves the transmission performance in the cases with and without delay constraints, by contrast to other benchmarks.

% Can use something like this to put references on a page
% by themselves when using endfloat and the captionsoff option.

%\bibliography{RIS-crossmedium}
%\bibliographystyle{IEEEtran}

\end{document}